# ZIGBEE BASED WIRELESS DATA ACQUISITION USING LABVIEW FOR IMPLEMENTING SMART DRIVING SKILL EVALUATION SYSTEM

Mohit John[1] and Arun Joseph[2]

Department of Applied Electronics and Instrumentation Engineering,
St. Joseph College of Engineering and Technology
Palai, Kerala, India
[1]mohitjohn83@gmail.com
[2]perhapsarun@gmail.com

***ABSTRACT***

*The Smart Driving Skill Evaluation (SDSE) System presented in this paper expedite the testing of candidates aspiring for a driving license in a more efficient and transparent manner, as compared to the present manual testing procedure existing in most parts of Asia and Pacific region. The manual test procedure is also subjected to multiple limitations like time consuming, costly and heavily controlled by the experience of examiner in conducting the test. This technological solution is developed by customizing 8051 controller based embedded system and LabVIEW based virtual instrument. The controller module senses the motion of the test vehicle on the test track referred to as zero rpm measurement and the LabVIEW based virtual instrument provides a Graphical User Interface for remote end monitoring of the sensors embedded on the test track. The proposed technological solution for the automation of existing manual test process enables the elimination of human intervention and improves the driving test accuracy while going paperless with Driving Skill Evaluation System. As a contribution to the society this technological solution can reduce the number of road accidents because most accidents results from lack of planning, anticipation and control which are highly dependent on driving skill.*

## 1. BACKGROUND

Despite continued efforts made by the different state governments in India, various international and national organizations continue to highlight the fatalities on the roads caused by inconsistent process of issuing driving licenses across India. The study conducted by the International Finance Corporation (IFC) indicates that the process of obtaining driving license in India is a distorted bureaucratic one. The independent survey conducted shows that close to 60 percent of license holders did not even have to take the driving license test and 54 percent of them were untrained todrive [1]. The study conducted by IFC also shows that the driving license is in that category of public services that involves corruption of a direct demand and supply of bribes between citizens and bureaucrats. The study also indicates that the corruption is focused on agents that work as intermediaries between the officials and citizens. This practice of agent-usage promotes corruption and subsequently results in higher payment for licenses, reduces driving test quality and this eventually results in unskilled drivers on road [1-2]. According to recent studies conducted in the US and UK have shown that about 95 percent of the road accidents are due to poor driving skills [3-4]. Hence the only solution for this problem is to implement an efficient, transparent and cost effective driver testing system.

## 2. EXISTING TEST APPROACH

In the present scenario, the candidates who have applied for driving license have to appear for a theoretical examination and a practical examination. The theoretical examination evaluates the candidate knowledge on different traffic signs, traffic regulation and also the basic understanding of simple safety check before using a vehicle. Different ways are adopted for the conduct of theoretical examination. These are oral examination, question paper or computer based examination. Theoretical examination is conducted before the practical examination [5]. A pass inthe theoretical examination is a prerequisite for the practical examination.

The practical examination comprises of two tests namely off-road test and on-road test. The off-road test is for examining the candidate's ability in controlling the vehicle. The on-road test is conducted in light traffic on normal road [3-5]. Normally, the on-road test is carried out after completing off-road test. The off-road test is performed on specially designed track, which are shown below.

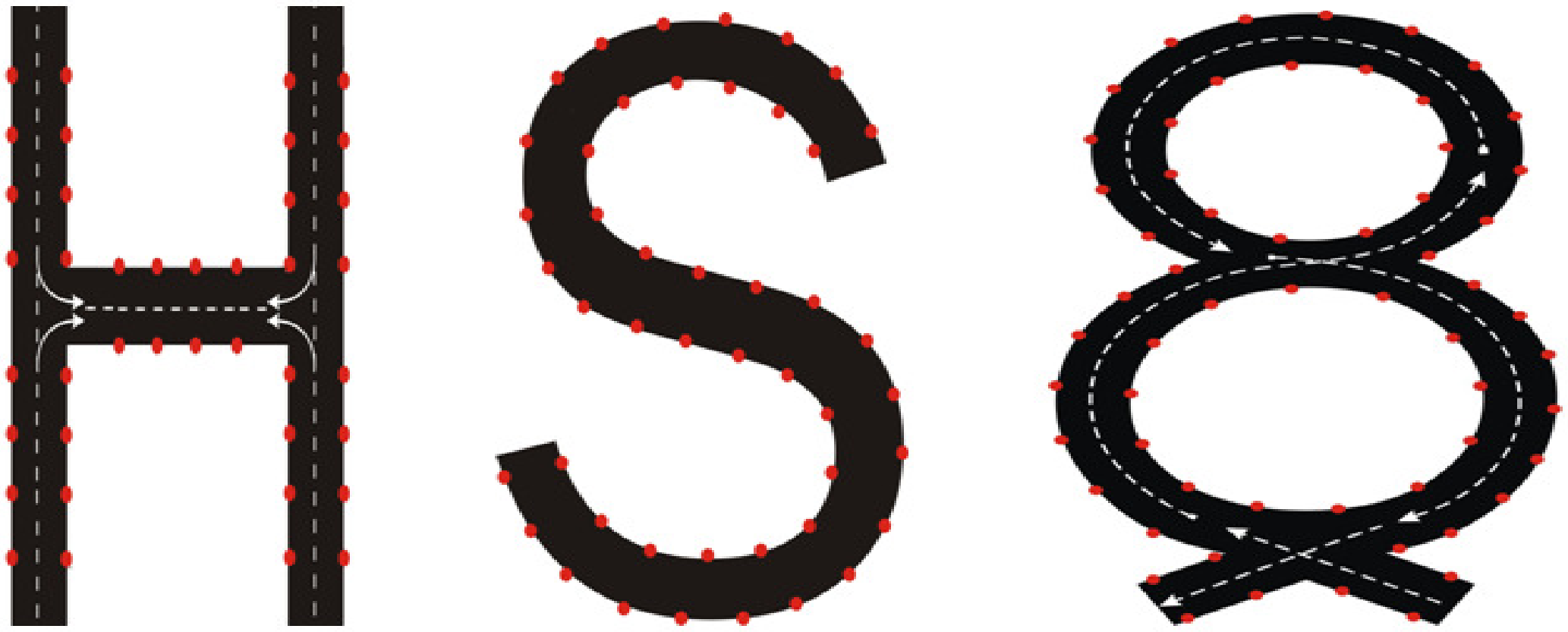

Figure 1. Types of off-road test track

The off-road test tracks are of three types – H, S and 8 shaped tracks. In India, the test track adopted for off-road test purpose varies from state to state. For example, the state of Kerala performs the off-road test using H- shaped track whereas the state of Karnataka performs off-road test using 8-shaped track. In most cases the evaluation of off-road test is done by human intervention. Hence the test result will be highly dependent on the subjective opinion of the examiner which is inconsistent. Hence to make the driving skill evaluation more transparent, consistent and efficient, the paper proposes a smart and automated system for driving skill evaluation.

## 3. SDSE SYSTEM SIGNIFICANCE AND FEATURES

The existing manual driving test procedure has got several limitations which demand the need for automated and smart driving skill testing system. The drawbacks of manual driving skill test are that the multiple examiners at locations evaluate the candidate's driving on subjective assessment, which leaves a scope for manipulation and negotiation [5]. Secondly the test results are paper documented, requiring storage space. These drawbacks can be eliminated with the SDSE system presented in this paper. The SDSE system also offers several advantages like total transparency, the consistency in test process, ensures selection only on merit and also prevents untrained drivers from getting into the system.

The SDSE is designed with a number of features like special tracks with sensors embedded in the track, real data transfer to central monitoring and result issuance system and on completion of the test, the results are printed and handed over to candidate immediately.

## 4. PROPOSED SDSE SYSTEM

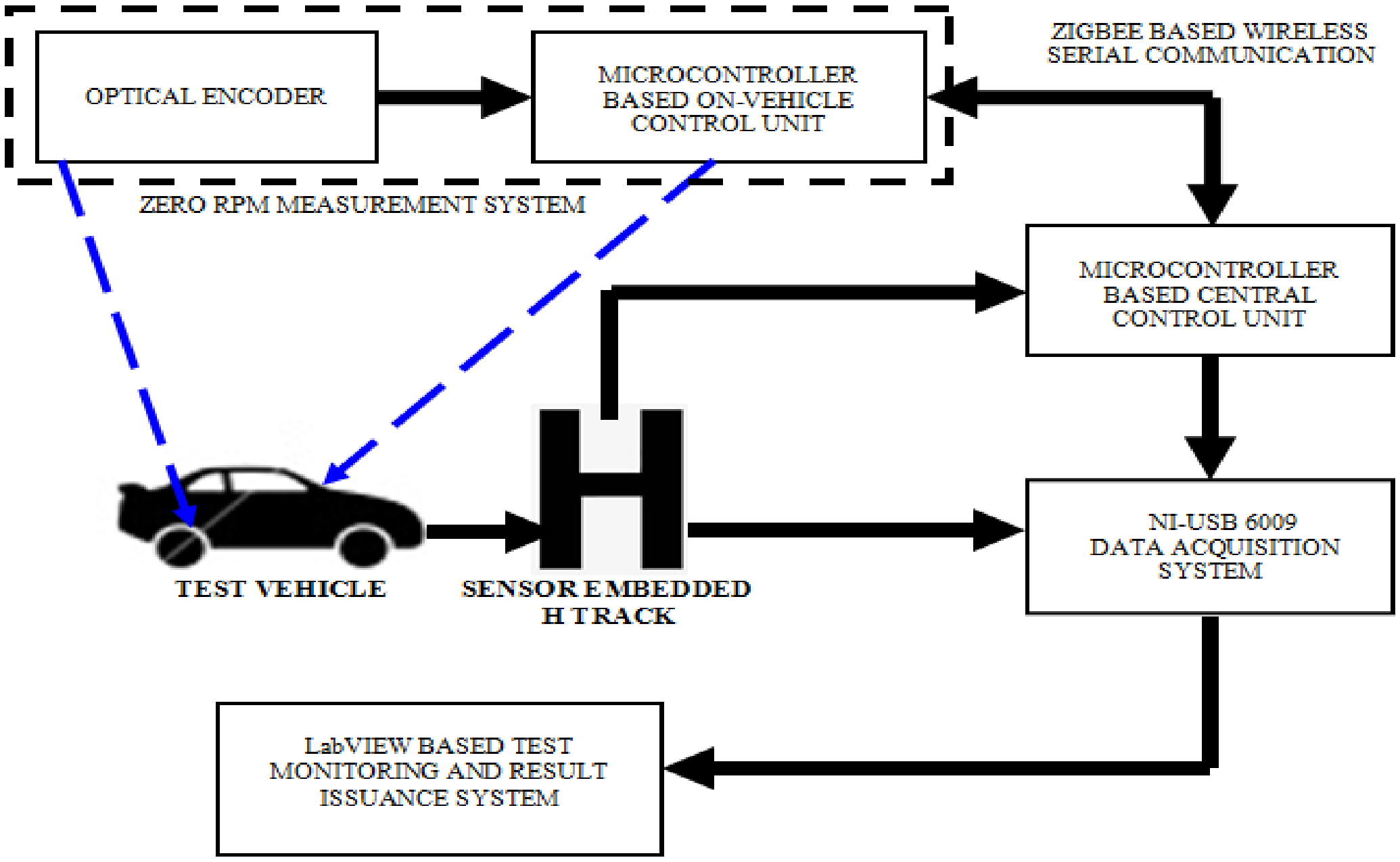

Figure 2. Simplified Representation of the SDSE System

The proposed system for evaluating the driving skill is implemented using H-shape based off-road test track which is shown in figure 2. In this system, initially the test candidate need to registration on the designated system with all supporting documents. The registration involves the submission of an e-application. The electronic application is developed on LabVIEW platform. The LabVIEW based application enables the system to store all the details during the test and generates a result card automatically, once the driving skill evaluation is completed. The SDSE system is designed in such a way that, a candidate will successfully complete the driving skill test only if the candidate completes the drive on the H track without trouncing the sensors embedded on the test track and also without crossing the line of intersection between sensors. In addition to the above conditions, the candidate needs to complete the H track without halting the test vehicle while driving on the H track. So to monitor these conditions, the SDSE system is automated with a LabVIEW based application to monitor the status of each sensor while the test is performed. For this the sensor embedded H track is interfaced to the LabVIEW based monitoring system using NI USB 6009 DAQ card. The test vehicle motion on the H track is sensed using 8051 microcontroller based zero rpm measurement system. The on-vehicle microcontroller based zero rpm measurement system is controlled by a 8051 microcontroller based central control unit. The communication between central control unit and zero RPM measurement system is established using ZigBee based wireless serial interface. The hardware used for ZigBee protocol based wireless serial communication is XBee Series 2 OEM RF module. The central control unit communicates with the LabVIEW based Test Monitoring and Result Issuance System through NI

USB 6009 DAQ card. In this application NI USB 6009 DAQ card digital I/O lines are used. Among the twelve digital I/O lines available in the DAQ card, eight I/O lines are used for interfacing eight sensor pairs (T1-R1 to T8-R8) embedded in the H track and one I/O line for interfacing with the central control unit.

## 4.1. Sensor Embedded H Track

The H track for the off-road test is monitored using the LabVIEW based remote system by embedding photo sensors on to the H track with the following measurement shown in figure 3.

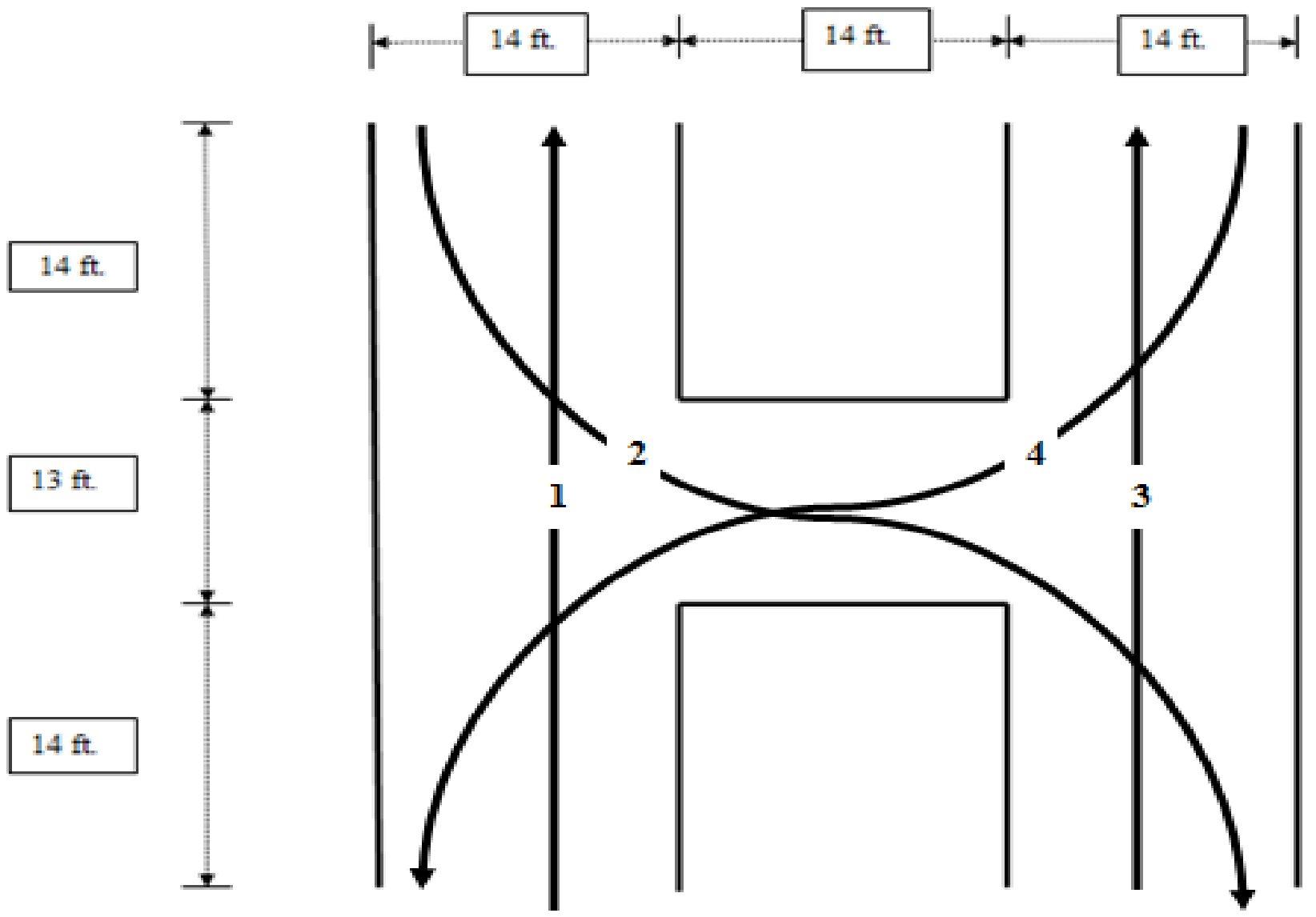

Figure 3. H track Layout

The drive on the H track is said to be complete only when the test candidate takes all the paths – 1, 2, 3 and 4 as shown in figure 3. The sensor embedded H track is implemented using Rif50 photo sensors. Rif50 photo sensors are referred to as synchronized, self-aligning photocells. Self-aligning feature makes the sensor pairs free from centering problem. Synchronism feature allows the installation of two pairs of very close photocells, without the problem of interference with one another. It is also unaffected by the interference of sunlight and has shockproof polycarbonate body.

| Parameter | Value |
|---|---|
| Dimension | 89x55x24 mm |
| Optical Range | 25m |
| Power Supply | 12 – 24 Vdc |
| Signal | Modulated infrared 833 Hz, $\lambda$ = 950nm |
| Relay Contact | 1 A max 30 Vdc |
| Absorption | TX 20mA RX 25mA |
| Operating Temperature | -20°C / +60°C |

Table 1. Rif50 Specification

The photo sensors are used as pairs where each pair consists of a transmitter and a receiver. The Rif50 photo sensors can be also referred to as a long range IR transmitter receiver pair. The Rif50 photo sensors are mounted on to yardstick for embedding on the H track. A yard stick mounted with the photo sensor is shown in figure 4.

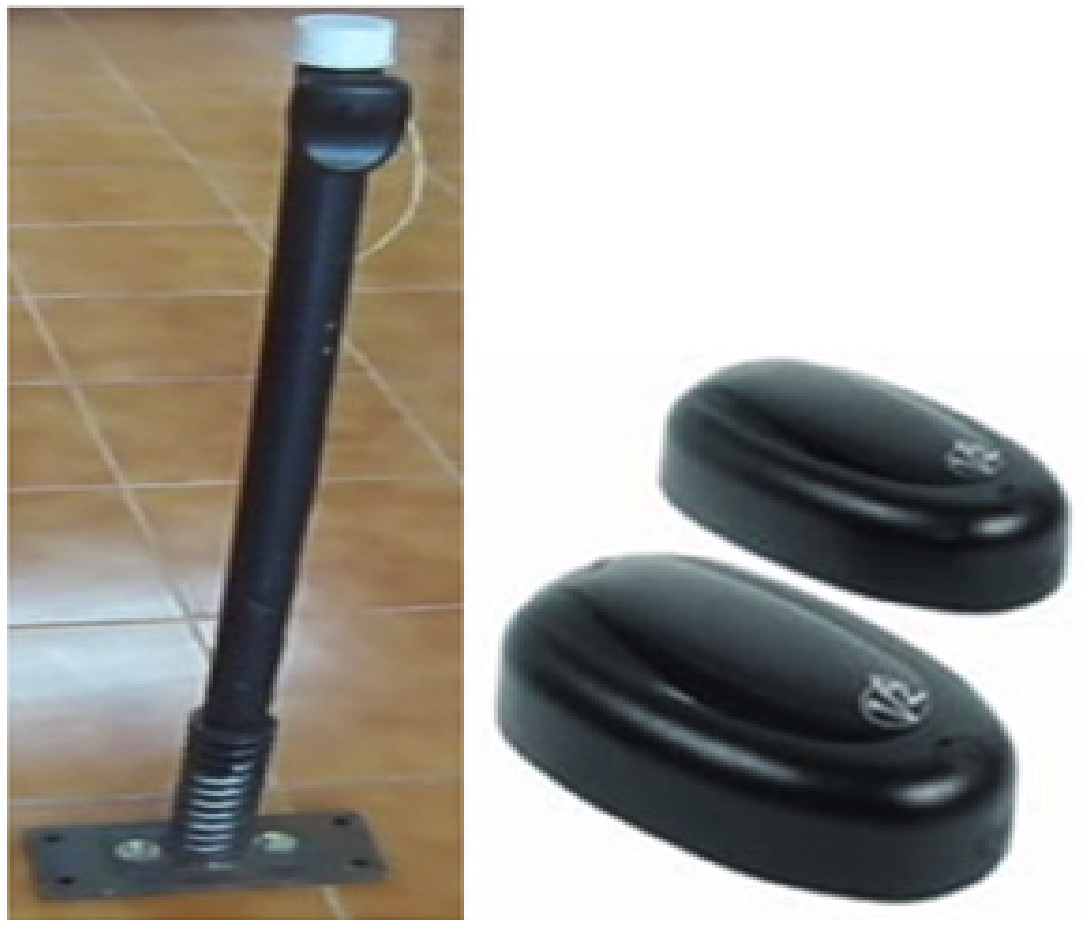

Figure 4. Rif50 photo sensor mounted yard stick

The layout for the photo sensor mounted yard stick embedded on the H track is shown in figure 5.

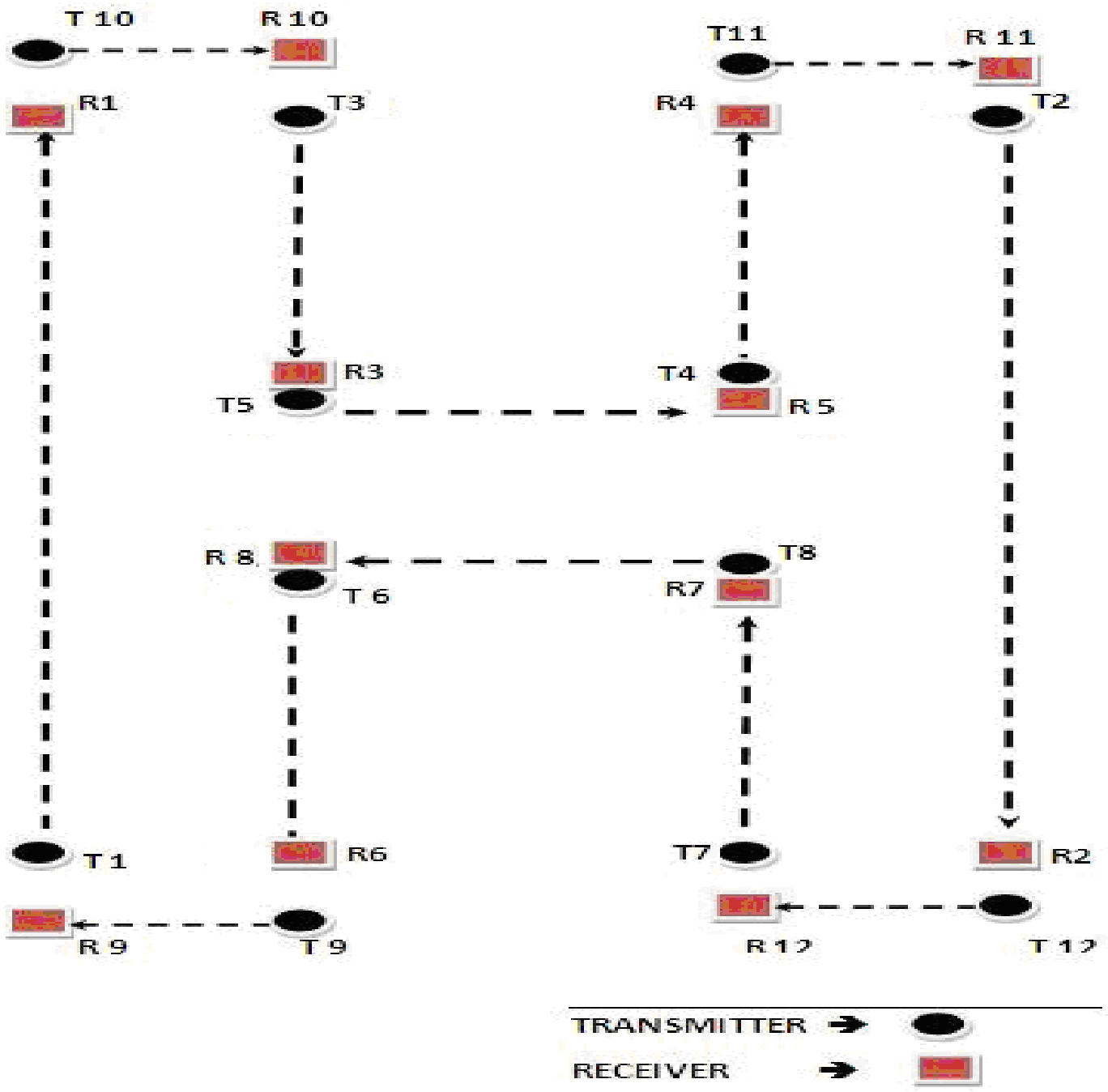

Figure 5. Layout for sensor embedded H track

Sensor embedded H track is implemented using twelve pairs of Rif50 photo sensors. Among the twelve sensor pairs, the eight sensor pairs labelled as T1-R1 to T8-R8 are for detecting the trouncing of sensor mounted yardstick or the crossing of line intersection between the sensor pairs. These sensors are interfaced to LabVIEW based Test Monitoring and Result Issuance system using NI USB 6009 DAQ card. The function of the four sensor pairs labelled as T9-R9 to T12-R12 is to enable or disable of zero rpm measurement system. These sensors are interfaced to the microcontroller based central control unit. When either of these sensor pairs makes a high to low transition, the microcontroller based central control unit enables or disables the microcontroller based on-vehicle control unit for monitoring the vehicle motion. The on-vehicle control unit for zero rpm measurement is designed to perform the measurement only when the vehicle is inside the sensor embedded H track. In other words, the on-vehicle control unit for zero rpm measurement is enabled only when the test vehicle is inside the H track and it is disabled when the test vehicle is outside the H track. The on-field implementation of the sensor embedded H track for test purpose is shown in figure 6

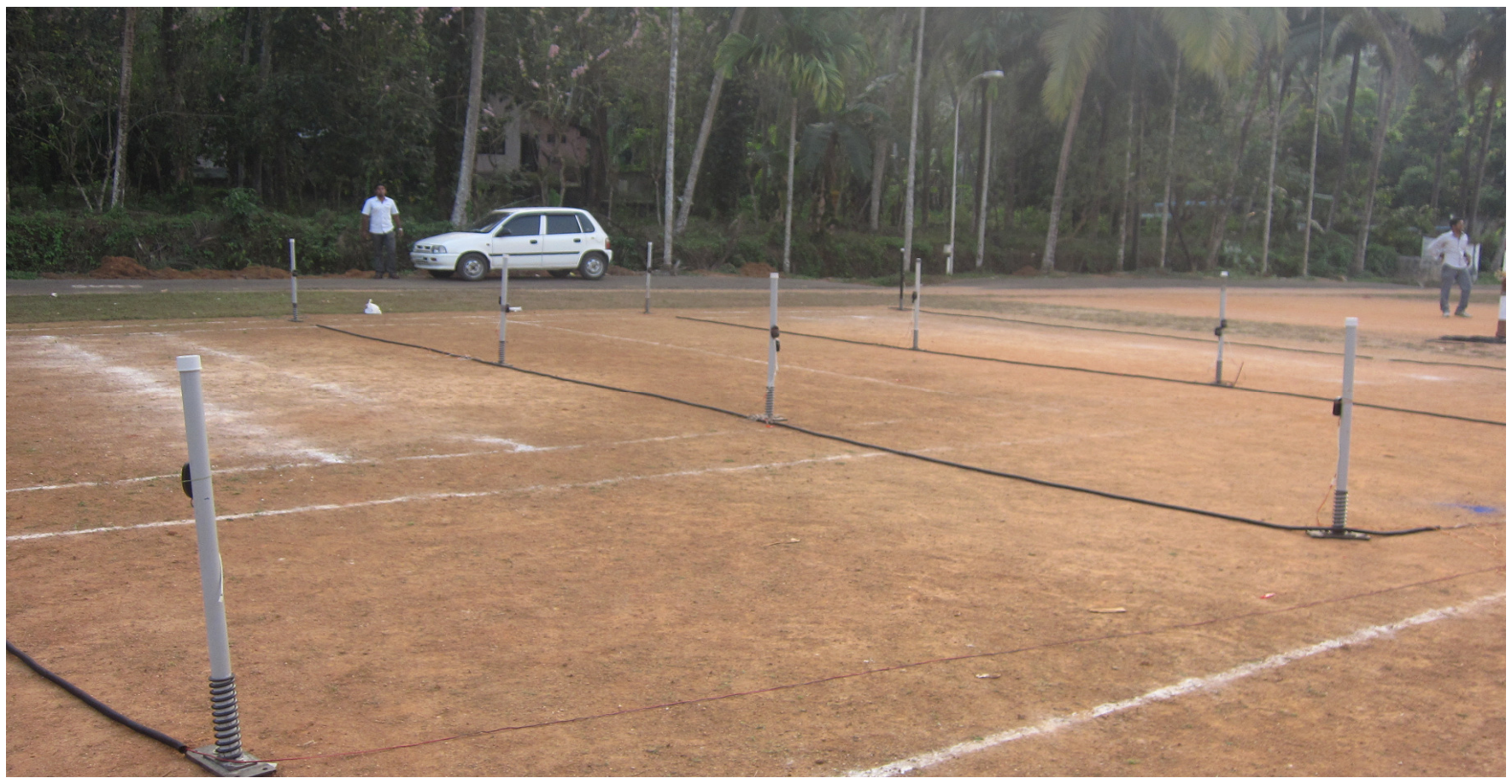

Figure 6. On-field sensor embedded H track

### 4.2. Data Acquisition System

The data acquisition system in the SDSE system is implemented using NI USB 6009 DAQ card. The DAQ card acquires the on-field H track photo sensor status to be monitored by the LabVIEW application. The NI USB 6009 based data acquisition system is shown in figure 7

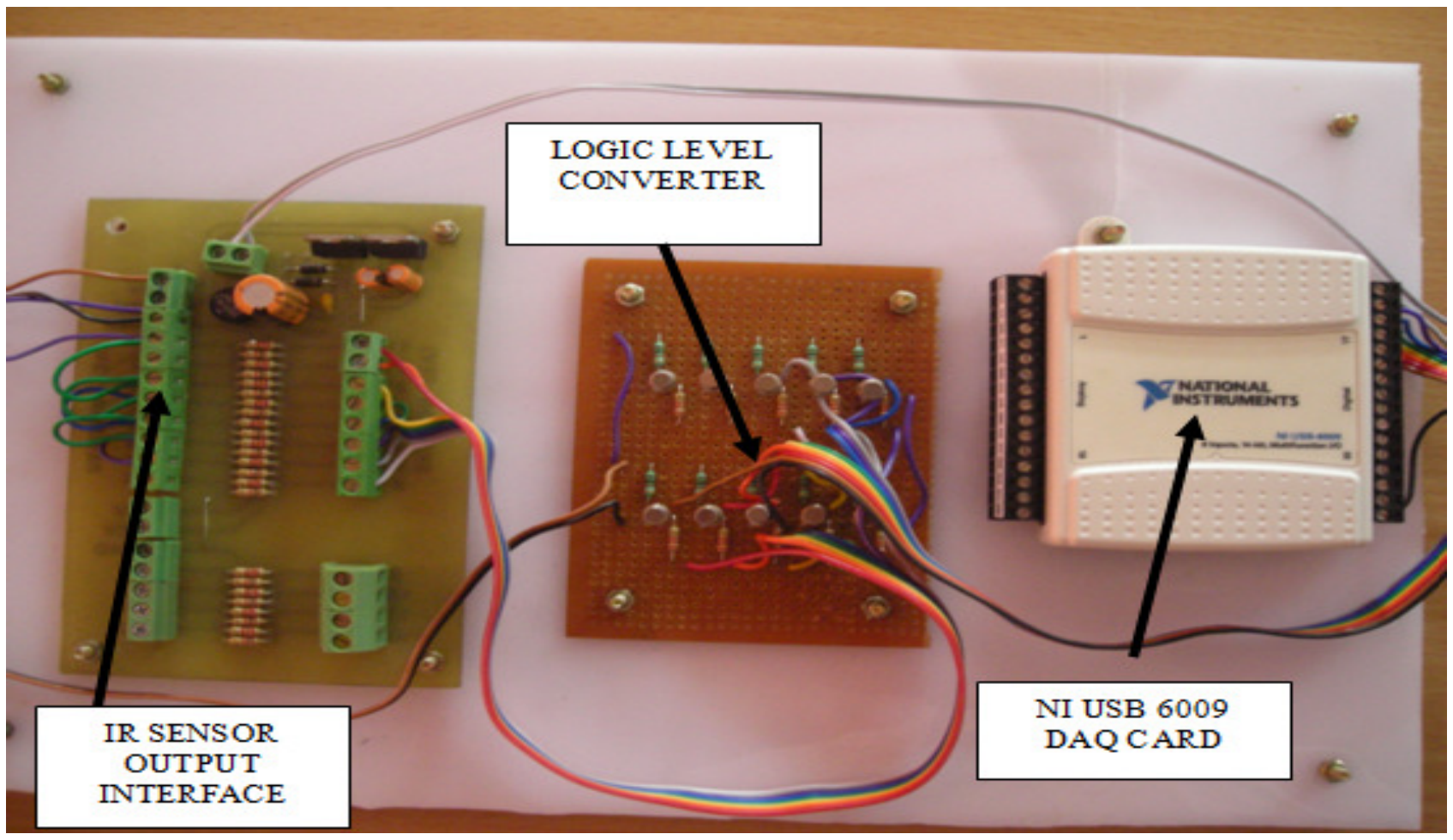

Figure 7. NI USB 6009 based Data Acquisition System

The Rif50 photo sensor pair gives two level digital outputs. The logic low level state is represented by 0 volt and the logic high state is defined by 12 volt. The logic level converter is used to make the output of photo sensors compatible with the voltage levels of digital I/O lines in NI USB 6009 DAQ card that supports TTL logic only. The information acquired by the DAQ card is interfaced to the LabVIEW based application on PC through USB interface.

### 4.3. ZigBee Based Wireless Data Acquisition

The wireless data acquisition system consists of an optical encoder, on-vehicle control unit, and a central control unit. The optical encoder assembly together with on-vehicle control unit is referred to as zero RPM measurement. The on-vehicle control unit and central control unit are 8051 microcontroller based system. The zero rpm measurement system consists is an optical encoder assembly as shown in figure 8.

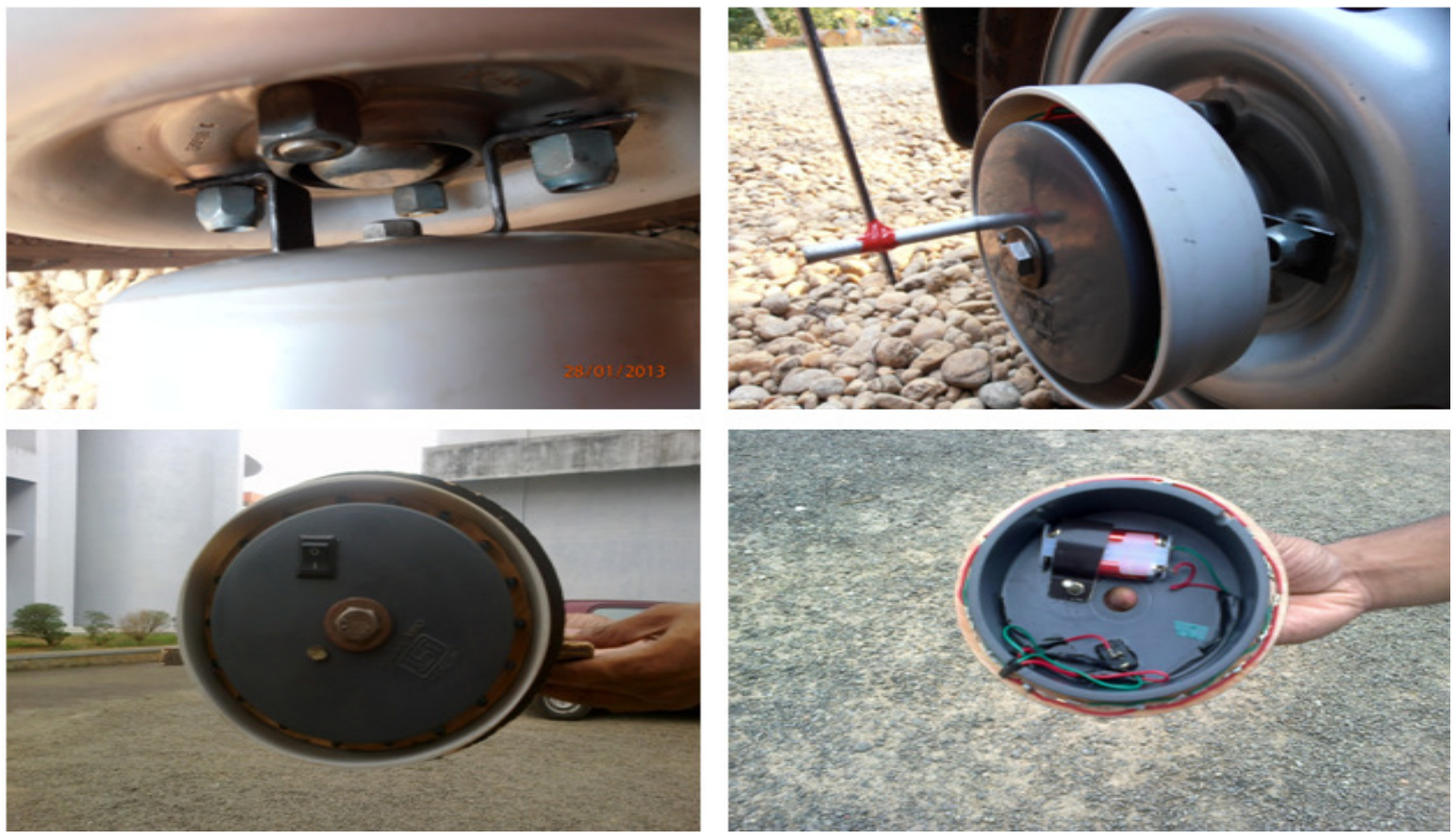

Figure 8. Hardware for Zero RPM Measurement System

The optical encoder consists of an array of light sources (LEDs) which are mounted on to one of the test vehicle wheel and a photo detector. The assembly comprising of array of LEDs is battery powered and it is mounted on to the test vehicle wheel as shown in figure 8. The array of light sources and the photo detector are mounted in such a way that their have the same line of action as shown in figure 9.

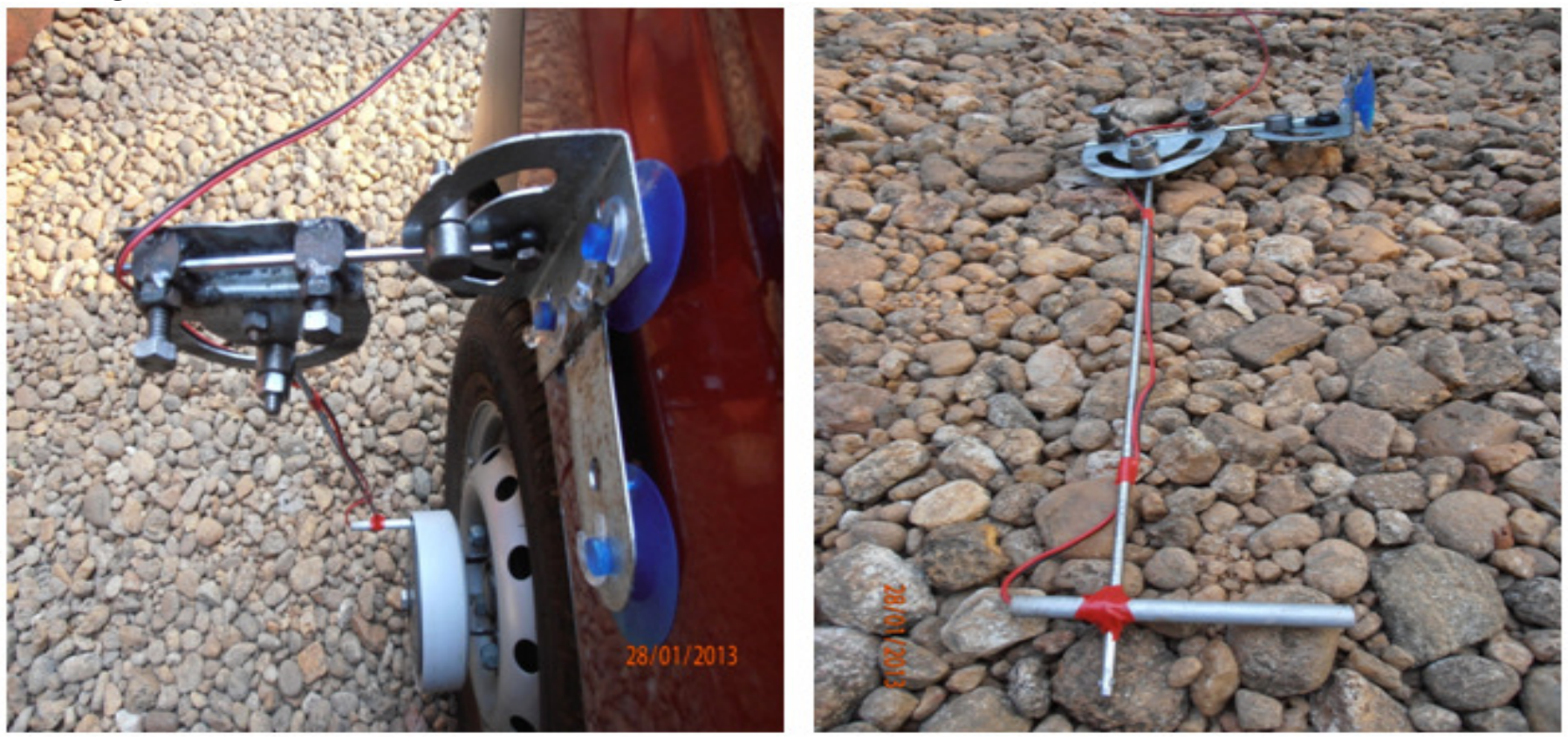

Figure 9. Zero RPM Measurement System mounting on Test Vehicle Wheel

The array of light sources rotates along with the test vehicle wheel. The optical encoder assembly output is a train of pulses which is proportional to the rotation of vehicle wheel. The pulse train output of the optical encoder assembly is shown in figure 10.

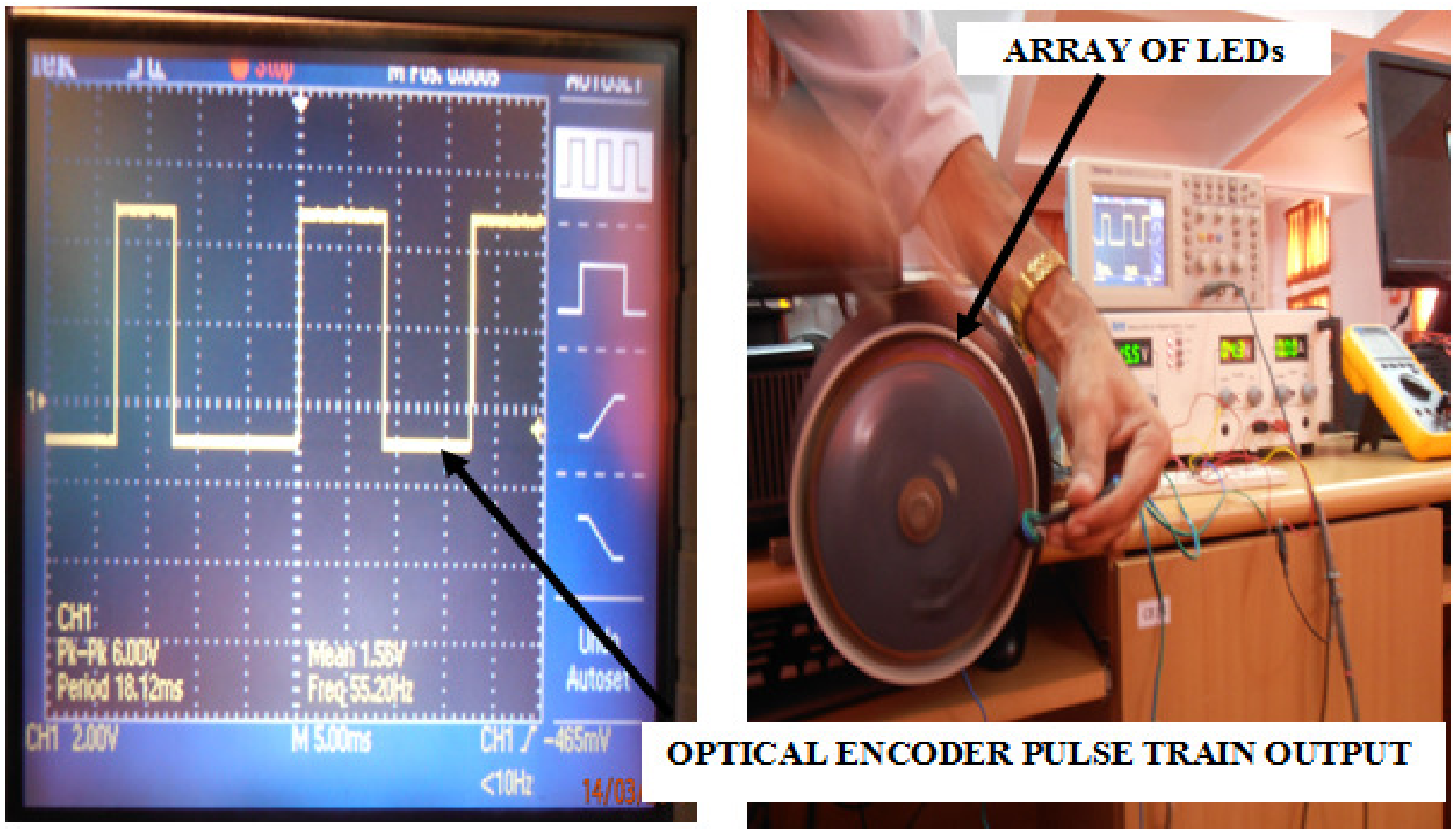

Figure 10. Test output of Zero RPM Measurement System

The pulse train output of optical encoder is interfaced as an external interrupt to the 8051 microcontroller based on-vehicle control unit. The hardware schematic of on-vehicle control unit is shown in figure 11.

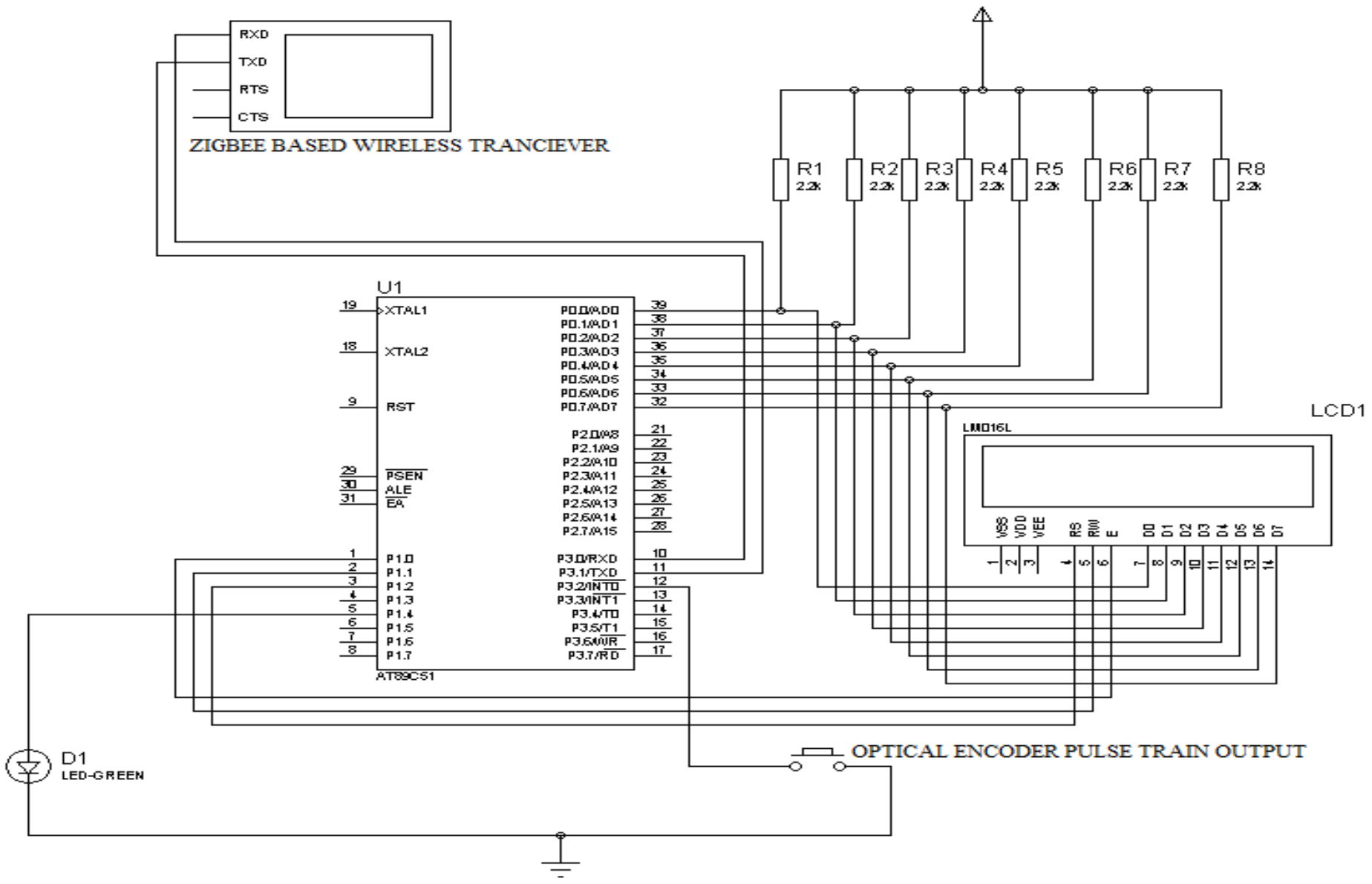

Figure 11. Hardware Schematic of on-vehicle control unit

The output of optical encoder assembly is connected to the external interrupt pin INT0. The 8051 microcontroller detects the pulse train output signal as a falling edge interrupt signal. When the on-vehicle control unit is enabled by the central control unit, the on-vehicle control unit monitors the number falling edges in pulse train output for every one second. The LCD interface in the on-vehicle control unit displays the enabling/disabling signal send by central control unit. The central control unit enables the on-vehicle control unit by sending the character ‘E’ and disables by sending the character ‘D’ through ZigBee based Wireless serial interface [6-9]. The information send by central control unit to enable/disable the on-vehicle control unit will be displayed on to the LCD interface available in the on-vehicle control unit as shown in figure 12.

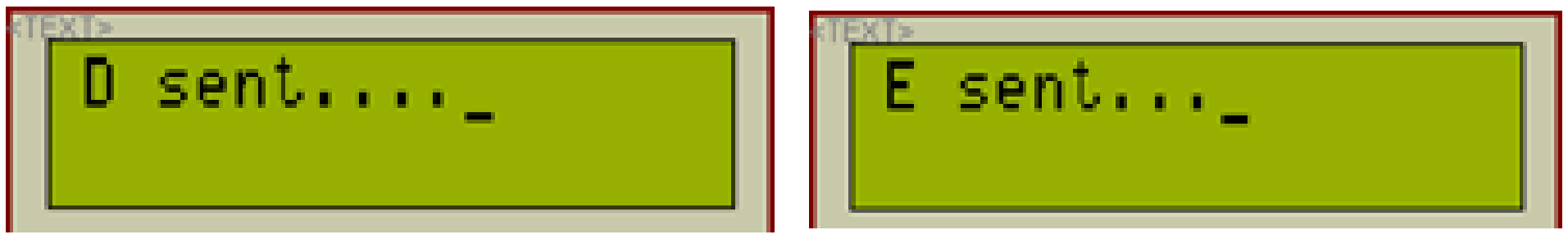

Figure 12. Information displayed on vehicle control unit LCD

The number of falling edges in the pulse train output of optical encoder is proportional to the rotational speed of the test vehicle wheel. Hence when the test vehicle wheel is stationary there will be no pulse train output from the optical encoder assembly and hence the number of falling edges detected will be zero. The on-vehicle control unit is designed to signal the central control unit through the ZigBee based wireless serial interface [8] when no falling edge is detected in one second. The hardware schematic of central control unit is shown in figure 13.

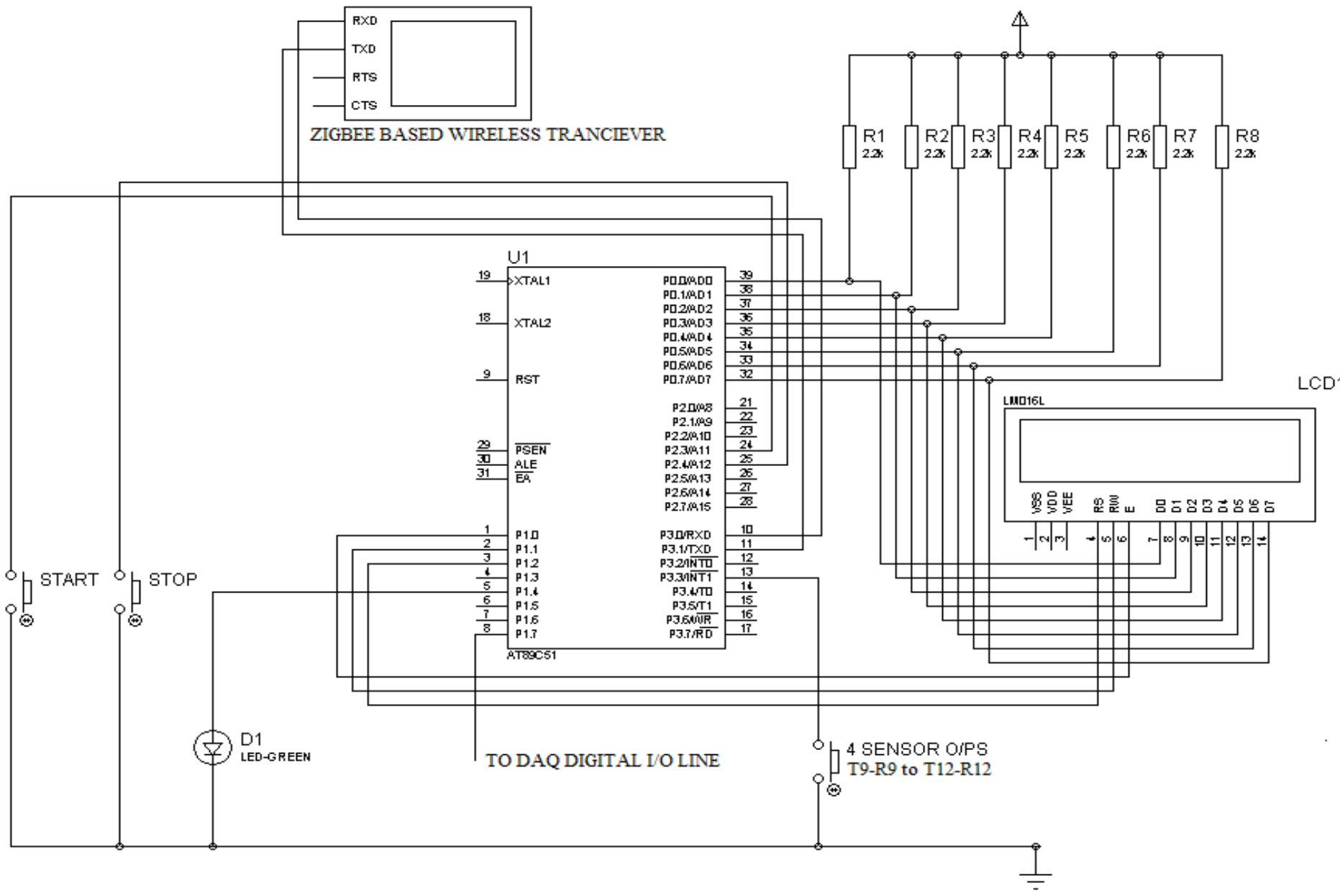

Figure 12. Hardware Schematic of central control unit

The communication between the central control unit and on-vehicle control unit is bidirectional wireless serial communication using ZigBee protocol [8-11]. The central control unit also monitors the four photo sensor pairs T9-R9 to T12-R12 to determine the presence of the test vehicle on the H track. The microcontroller based system determines the presence of the test vehicle by counting the number of High to Low transition output signal of photo sensor pairs (T9-R9 to T12-R12). These four photo sensor pairs are connected to INT1 pin of 8051 microcontroller using a logical OR gate circuit. The central control unit communicates with the on-vehicle control unit for enabling or disabling the zero RPM measurement. Zero RPM measurement is enabled only when the test vehicle is inside the H track. From figures 3 and 4, it can be said that initially the test vehicle takes the path1 by crossing the line of action of the photo sensor pair T9-R9. Now this photo sensor pair output makes a High to Low transition, which will be taken as a falling edge interrupt signal – INT1 by the 8051 microcontroller. This High to Low transition output in the T9-R9 photo sensor pair will be count one for the microcontroller. At the end of path1, the test vehicle crosses the line of action of T10-R10 photo sensor pair. Now this photo sensor pair output makes High to Low transition, which will be taken as a falling edge interrupt signal – INT1 by the microcontroller and the internal count becomes two. Crossing the line of action of the photo sensor pairs – T9-R9 and T10-R10, the test vehicle is said to complete path1. Now the test vehicle takes the path2, by first crossing the line of action of the photo sensor pair T10-R10, thereby producing a High to Low transition which will be taken as another falling edge interrupt signal and now the internal count becomes three. The test vehicle completes path2, by crossing the line of action of the photo sensor pair T12 – R12 and now the count becomes four. Once the path2 is completed, the test vehicle starts with path3 by crossing the line of action of photo sensor pair T12-R12. Now the internal count of the microcontroller becomes five. The path3 is completed by crossing the line of action of photo sensor pair T11-R11 and thereby changing the internal count value to six. Finally the test vehicle takes the path4 by first crossing the line of action of T11-R11, which changes the count value to seven. The path4 is completed by the test vehicle by crossing the line of action of photo sensor pair T12-R12 and thereby finally making the count value to eight. From the above algorithm description, it is clear that when the count value is one, three, five and seven, the vehicle is inside the H track. For these count values the central control unit enables zero RPM measurement system by sending serially the character E to the on-vehicle control unit. For the count values two, four, six, eight the test vehicle is said to be outside the H track. Hence for these count values the central control unit disables the zero RPM measurement system by sending serially the character D to the on-vehicle control unit. During the enabled state of the on-vehicle control unit, if a zero falling edge signal is detected for more than one second, then the on-vehicle control unit signals the central unit and the same will be displayed on to the LCD interface of the central control unit as shown in figure 13.

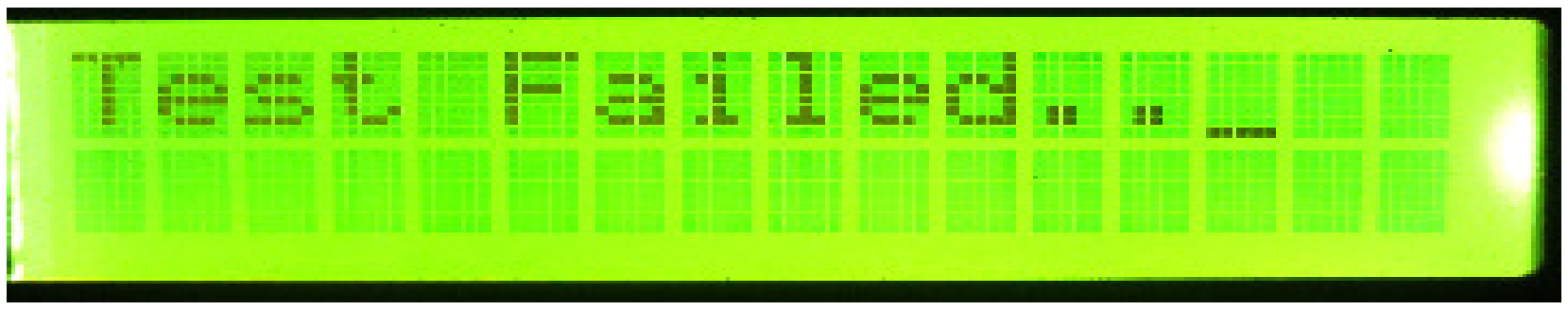

Figure 13. Information displayed on central control unit LCD

The central control unit signals the test failed condition to the LabVIEW based Test Monitoring and Result Issuance System through NI USB DAQ card. Finally the LabVIEW based GUI informs the test status to the operator.

## 4.4. LabVIEW Based Test Monitor and Result Issuance System

The test monitoring and result issuance application is developed using LabVIEW software. LabVIEW (short for Laboratory Virtual Instrument Engineering Workbench) is a system design platform and development environment for a visual programming language from Instruments. The graphical language is named "Z" (not to be confused with G-code) [12-13]. Originally released for the Apple Macintosh in 1986, LabVIEW is commonly used for data acquisition, instrument control, and industrial automation on a variety of platforms including Microsoft Windows, various versions of UNIX, Linux, and Mac OS X. The programming language used in LabVIEW, also referred to as G, is a dataflow programming language. Execution is determined by the structure of a graphical block diagram (the LV-source code) on which the programmer connects different function-nodes by drawing wires [16]. These wires propagate variables and any node can execute as soon as all its input data become available. Since this might be the case for multiple nodes simultaneously, G is inherently capable of parallel execution. The LabVIEW based test monitoring and result issuance application consist of two GUI. The first GUI is designed for monitoring the status of photo sensor on the H track. This GUI monitors whether the sensors are aligned or misaligned. The second LabVIEW based GUI is the e-application which acquires test candidate's details for generating the test result card soon after the test. The LabVIEW based GUI design for monitoring the on-field H track sensor is shown in figure 14.

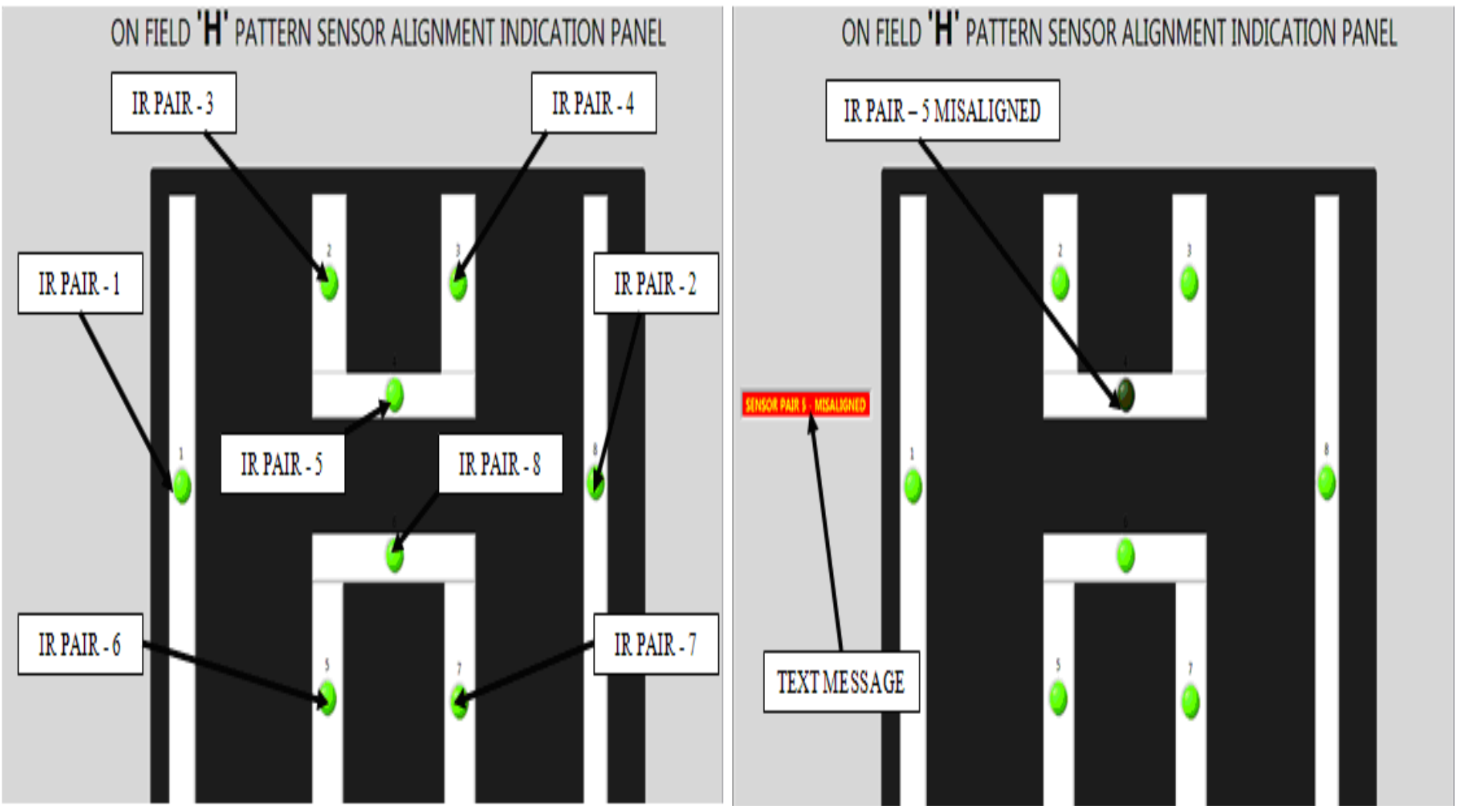

Figure 14. LabVIEW based GUI design for H track monitoring

Here the status of all the eight IR transmitter-receiver pairs on the H-track is mapped on to the LabVIEW based GUI shown in figure 14. The mapping of on-field IR transmitter-receiver pairs is done in accordance with the sensor embedded H track layout as shown in figure 5. In the LabVIEW based GUI monitoring the sensor pair alignment, the sensor pair alignment is animated using eight LED indicators. The ON status of animated LED indicates that the sensor is aligned and OFF status indicates that the sensor is misaligned. An example for such an instance is shown in figure 14, here the animated LED corresponding to sensor pair 5 is OFF which implies that IR sensor pair 5 on the H-track is misaligned. In addition to this, the LabVIEW based GUI informs the operator by displaying a text message regarding the misaligned sensor pair as shown in figure 14. This feature reduces the difficulty in indentifying the misaligned sensor pair embedded on to the on-field H track.

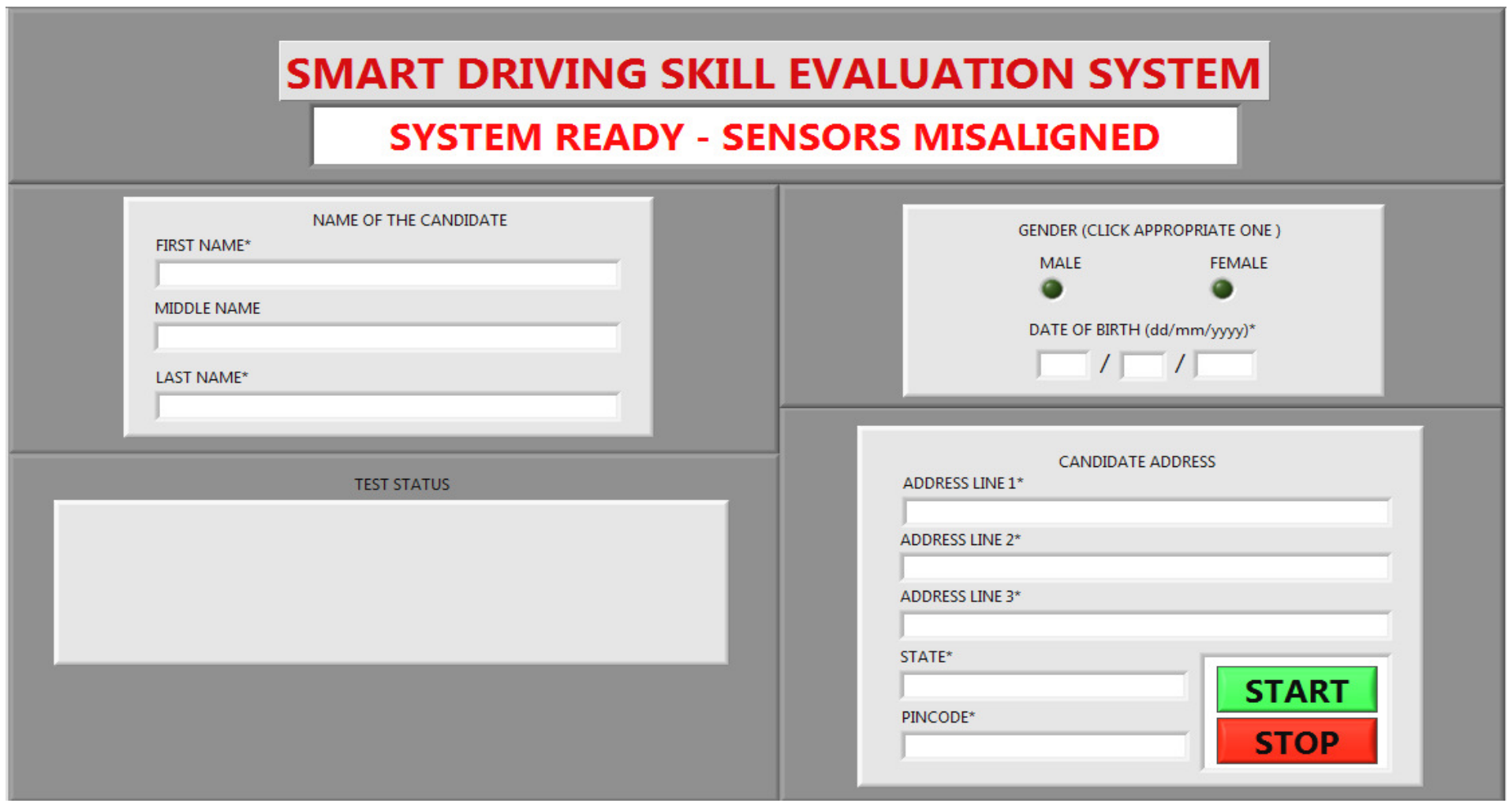

Figure 15. LabVIEW based GUI design for e-application

The LabVIEW based GUI for e-application accept the test candidate's following personal details – first name, middle name, last name, address for communication, date of birth and the gender. The GUI for e-application designed on LabVIEW platform is shown in figure 15. The GUI for e-application accepts test candidate information on pressing the START button only if all the photo sensors pairs on the H track are properly positioned. When the sensor pairs on the H track are not properly positioned, the GUI will show the message: SYSTEM READY – SENSORS MISALIGNED as show in figure 15. When the entire sensor pairs on the H track are properly aligned, the LabVIEW based GUI will show the message: SYSTEM ACTIVE – FILL IN CANDIDATE DETAILS as shown in figure 16.

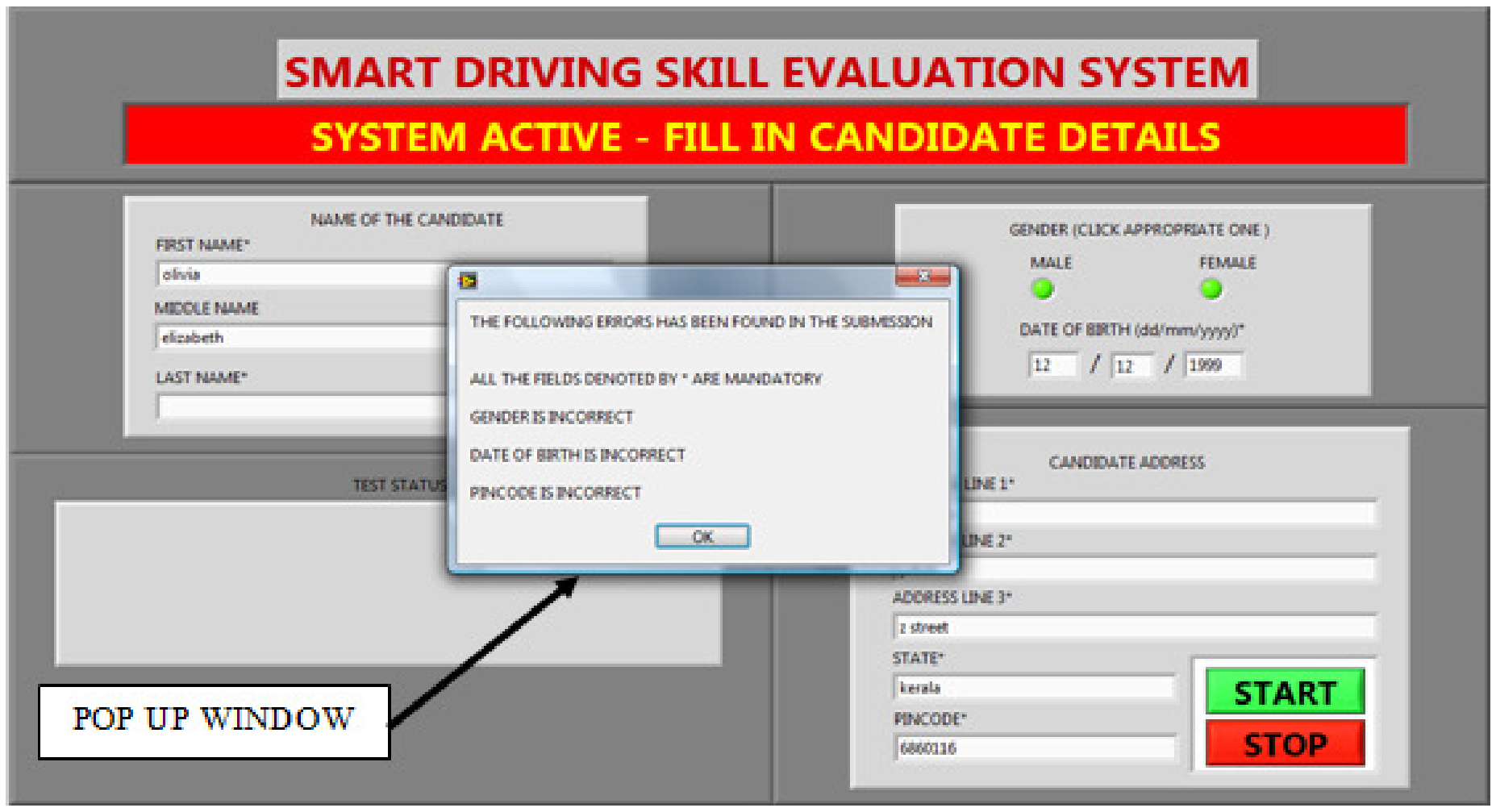

Figure 16. Error(s) in submitted LabVIEW based GUI design

The LabVIEW based GUI is designed to identify error(s) in the submitted e-application. The error(s) that are identified by the LabVIEW based e-application are blank data fields which are mandatory, incorrect gender, incorrect date of birth and incorrect pin code. On submitting an e-application with these incorrect information by pressing START button will generate a pop up window as shown in figure 16. The pop up window will inform the operator about all the incorrect information in the submitted e-application.

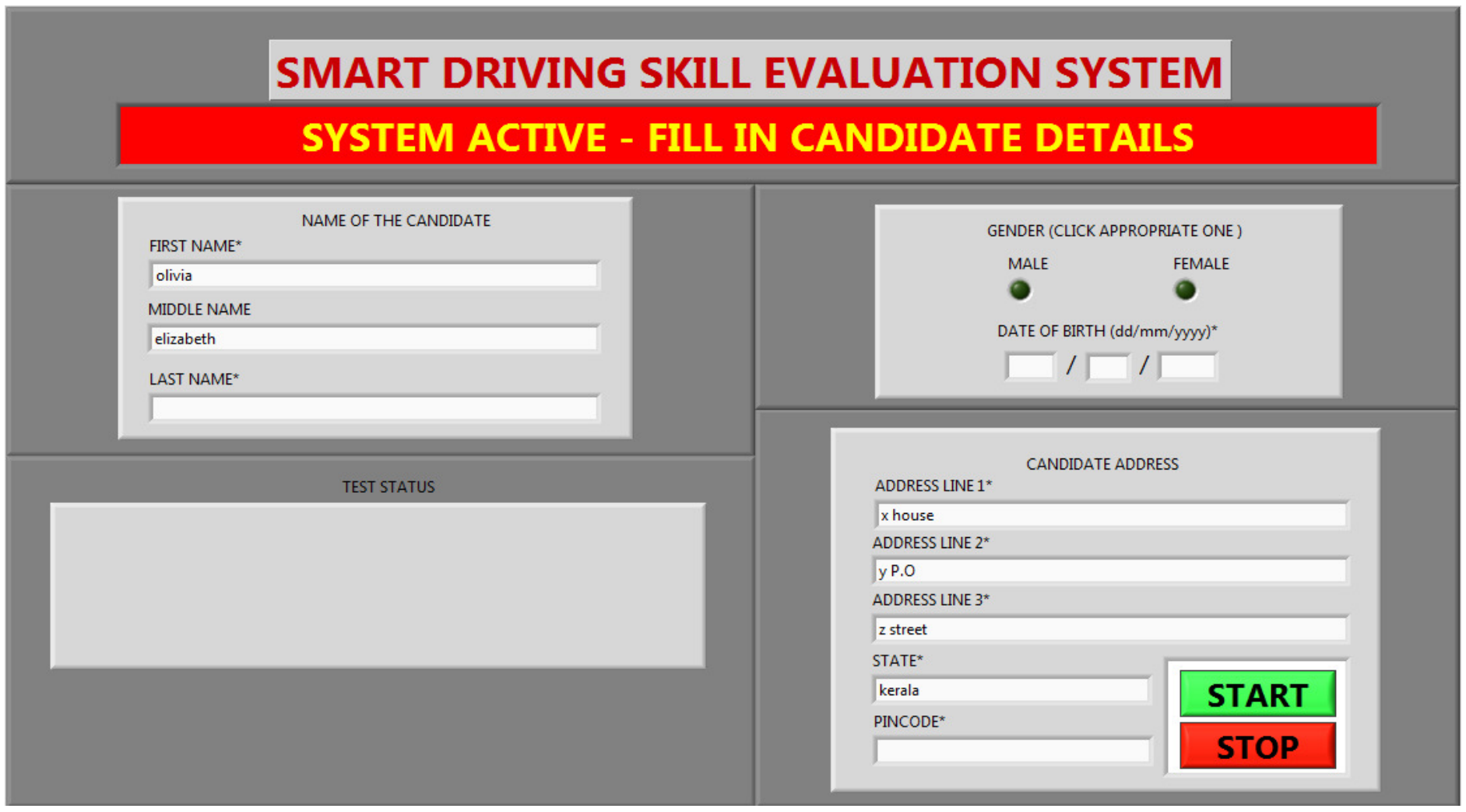

Figure 17. Incorrect fields cleared on pressing OK button

On pressing the OK button in the pop up window, all the incorrect fields in the LabVIEW based GUI will be cleared while retaining the correct fields in the e-application which is shown in figure17. To continue with the driving skill test process, the e-application should be submitted successfully. If all the information fields in the LabVIEW based e-application is completed correctly, then on pressing the START button another pop up window appears as shown in figure 18.

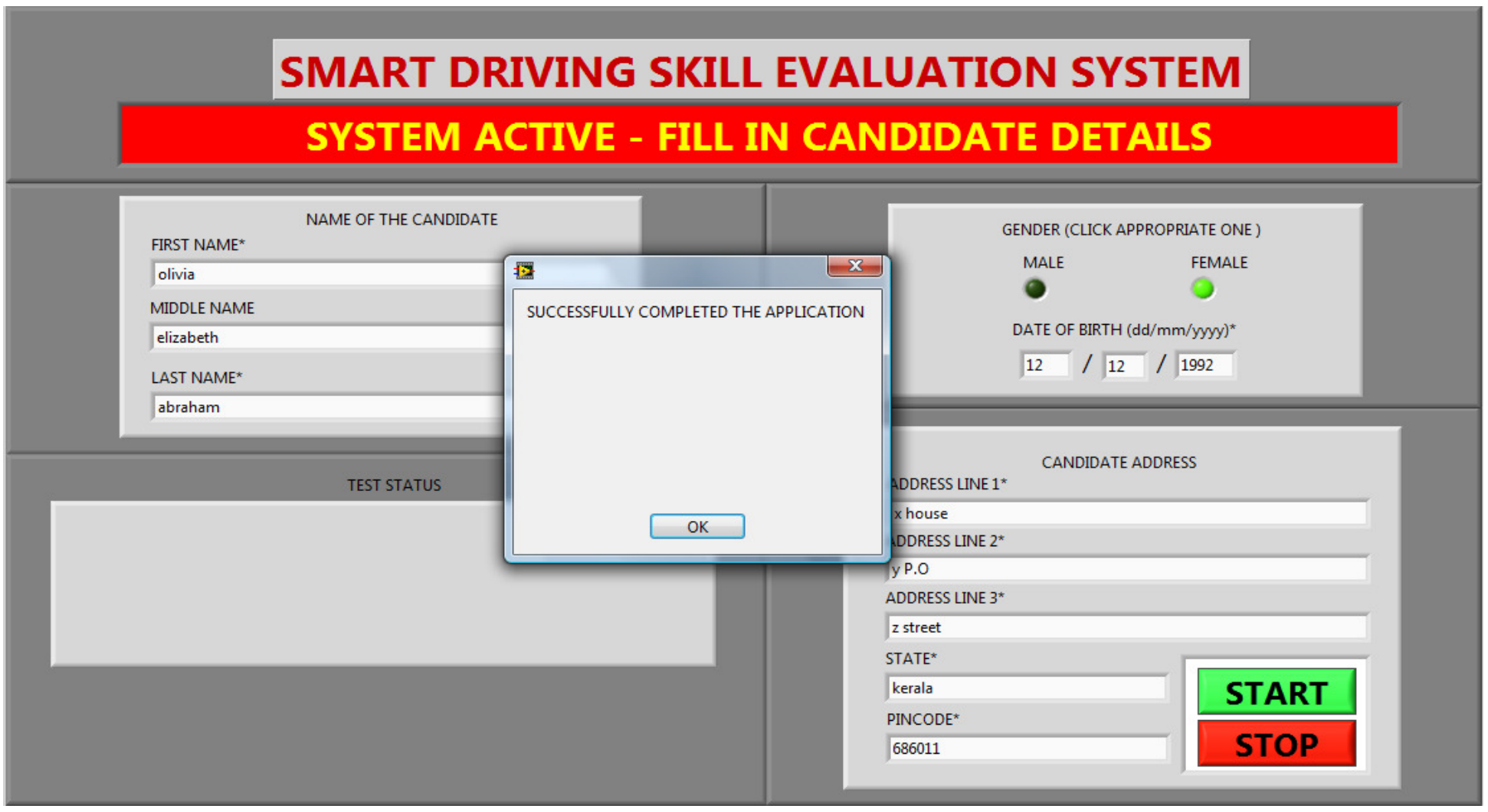

Figure 18. Successfully completed LabVIEW based e-application

With the successfully completed LabVIEW based e-application the user has to acknowledge by pressing the OK button on the pop up window as shown in figure 18. On pressing the OK button, the system will give an option for the user to cancel or continue further with the driving skill evaluation process as shown in figure 19. To continue with the test process the user has to press the SUBMIT button and to terminate the test process press the CANCEL button. When the CANCEL button is pressed, all the fields in the e-application will be cleared for the registration of next test candidate.

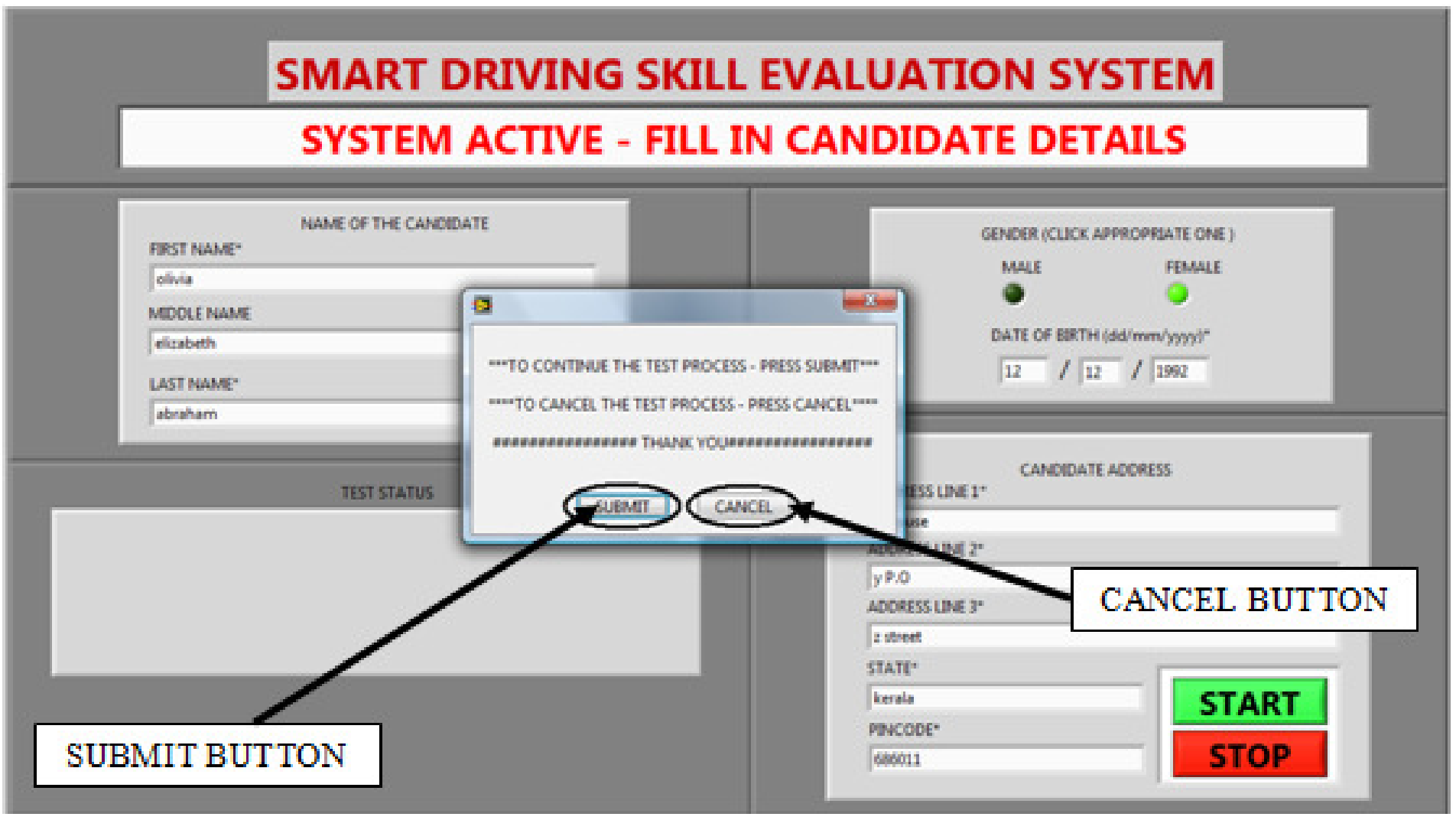

Figure 19. Submit and Cancel option with LabVIEW based e-application

On pressing the SUBMIT button, the information regarding the test candidate will be recorded with the system for the generation of test result card, after which the candidate can move to the test vehicle to drive on sensor embedded H-track. During the driving on the sensor embedded H track, if the test candidate trounces any of the sensors or crosses the line of intersection between the sensors or even stop the test vehicle while the drive on the H track is progressing then the LabVIEW based GUI will display: TEST FAILED as shown in figure 20.

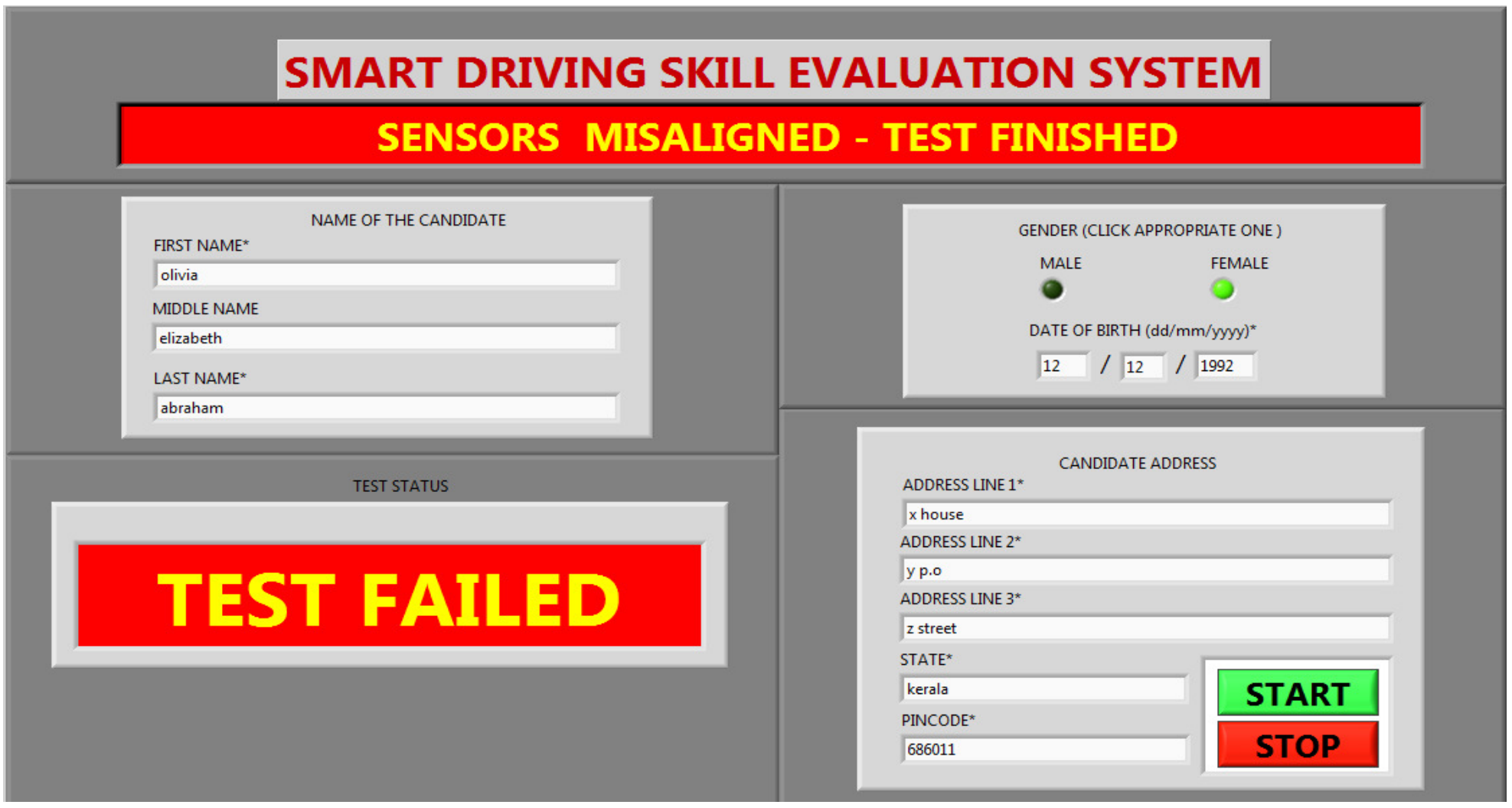

Figure 20. LabVIEW based GUI with TEST FAILED status

In addition to the failed test status, the LabVIEW based e-application also indicates the reason for failure. If the test failure is due to trouncing of sensors or crossing the line of intersection between the sensors, then the following message will be displayed: SENSORS MISALIGNED – TEST

FINISHED. If the test failure is due to vehicle stoppage then the following message will be displayed: VEHICLE HALT – TEST FINISHED. Now if the test candidate has successfully completed the drive on the sensor embedded H track, then the STOP button has to be pressed. On pressing the STOP button, then the test status will be displayed as TEST PASSED on the LabVIEW based GUI as shown in figure 21.

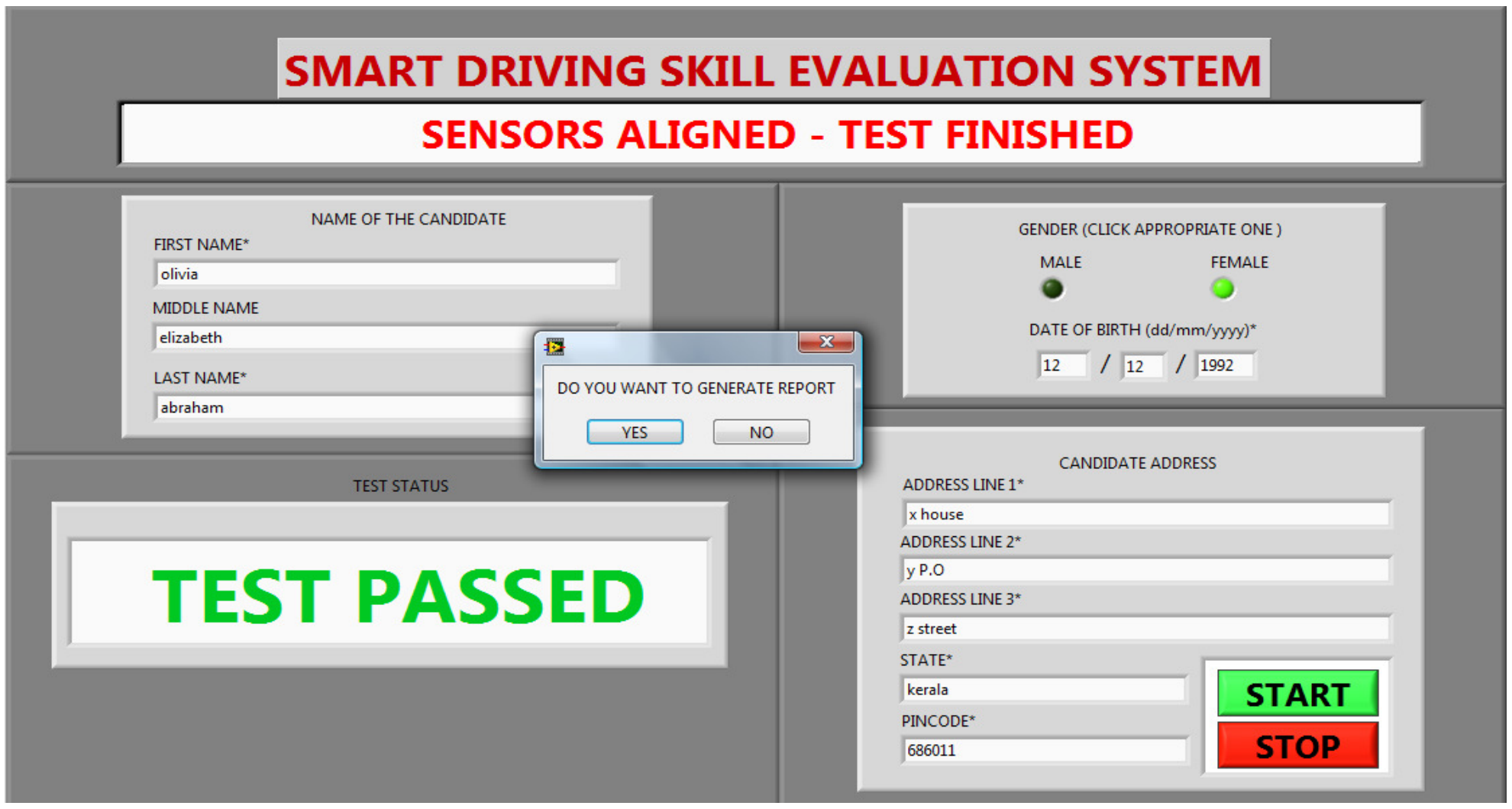

Figure 21. LabVIEW based GUI with TEST PASSED status

As soon as the test status is displayed on the LabVIEW based GUI, another pop window appears asking for test report generation. If the test result card needs to be generated then the YES button has to be pressed and if not then the STOP button has to be pressed. On pressing the YES button, the system will ask the operator for saving the test report as a MS word document with file extension .doc and the location where the file has to be saved as shown in figure 22.

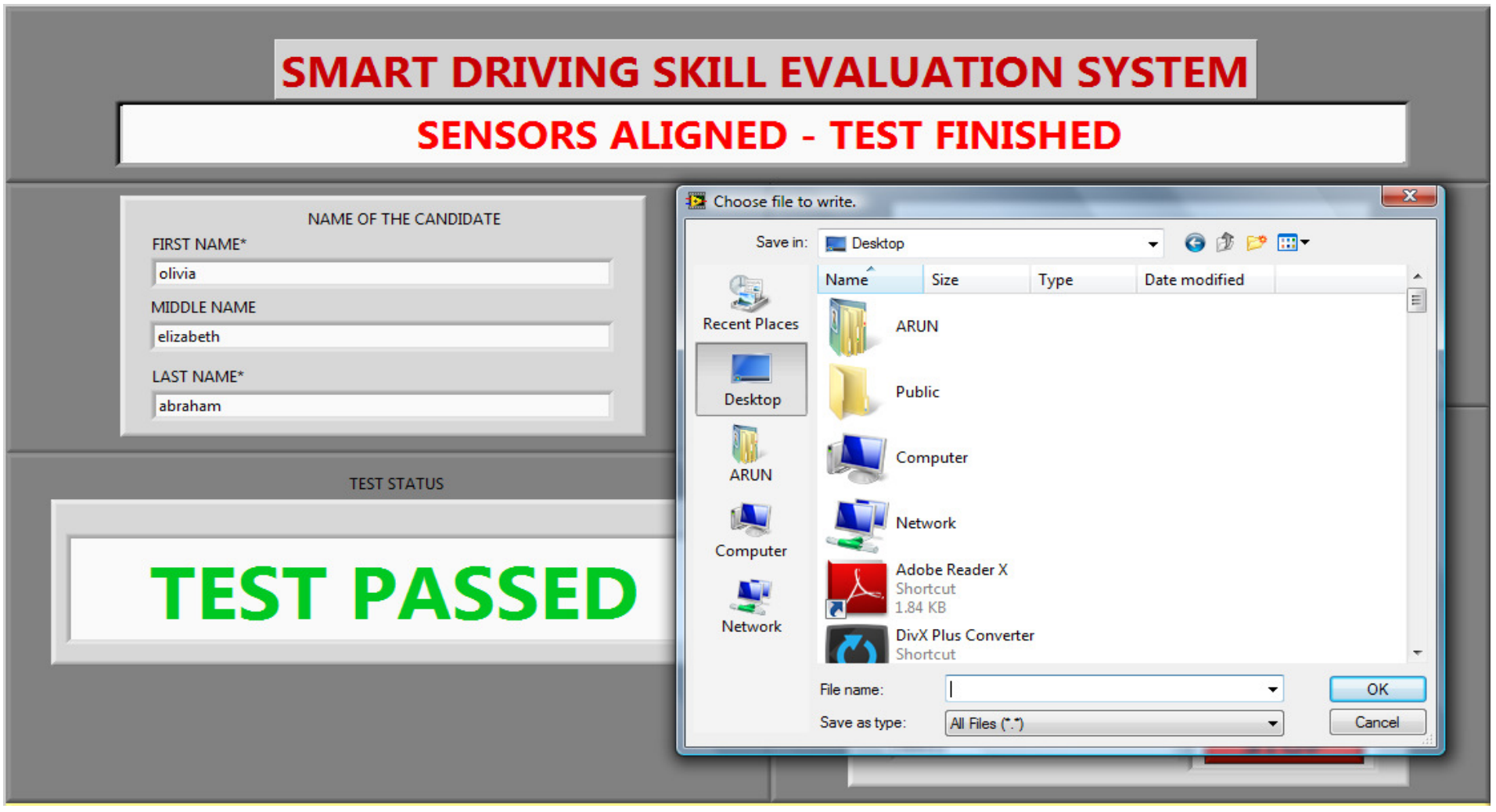

Figure 22. Saving the Test Report

Once the test report is saved the system automatically initializes and makes the LabVIEW based e-application ready for the registration of next test candidate. If the NO button is pressed then the system will not save the test report and instead directly makes the LabVIEW based e-application ready for the next test candidate. During the initialization it clears all the fields of e-application and checks whether all the sensors in the automated H-field is aligned or not. The screen shot of saved driving test result card is shown in figure 23. The test result card shows the date and time of the test process, the test status whether passed or failed and finally shows all the personal details of the test candidate given in the e-application during the registration process.

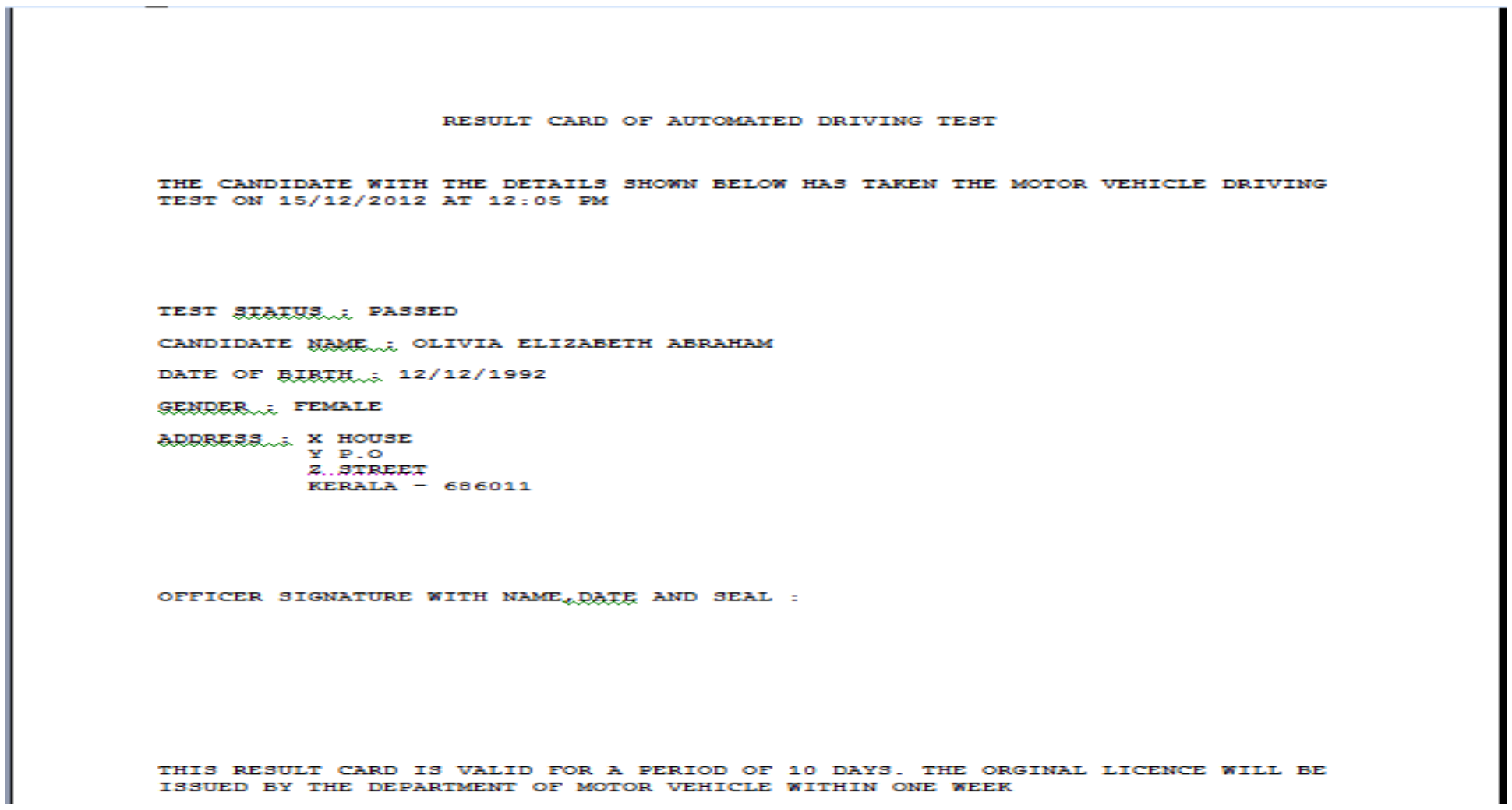

RESULT CARD OF AUTOMATED DRIVING TEST

THE CANDIDATE WITH THE DETAILS SHOWN BELOW HAS TAKEN THE MOTOR VEHICLE DRIVING TEST ON 15/12/2012 AT 12:05 PM

TEST STATUS : PASSED

CANDIDATE NAME : OLIVIA ELIZABETH ABRAHAM

DATE OF BIRTH : 12/12/1992

GENDER : FEMALE

ADDRESS : X HOUSE
Y P.O
Z.STREET
KERALA - 686011

OFFICER SIGNATURE WITH NAME,DATE AND SEAL :

THIS RESULT CARD IS VALID FOR A PERIOD OF 10 DAYS. THE ORGINAL LICENCE WILL BE ISSUED BY THE DEPARTMENT OF MOTOR VEHICLE WITHIN ONE WEEK

Figure 23. Screen shot of LabVIEW generated test result card

The on-vehicle control unit performing zero rpm measurement communicates serially with the LabVIEW based system. The LabVIEW based application establishes serial communications using National Instrument-Virtual Instrument Software Architecture (NI-VISA) read and write functions defined in LabVIEW 8.6 [14-16]. The implementations of serial read and write operation using NI-VISA functions is shown in figure 24.

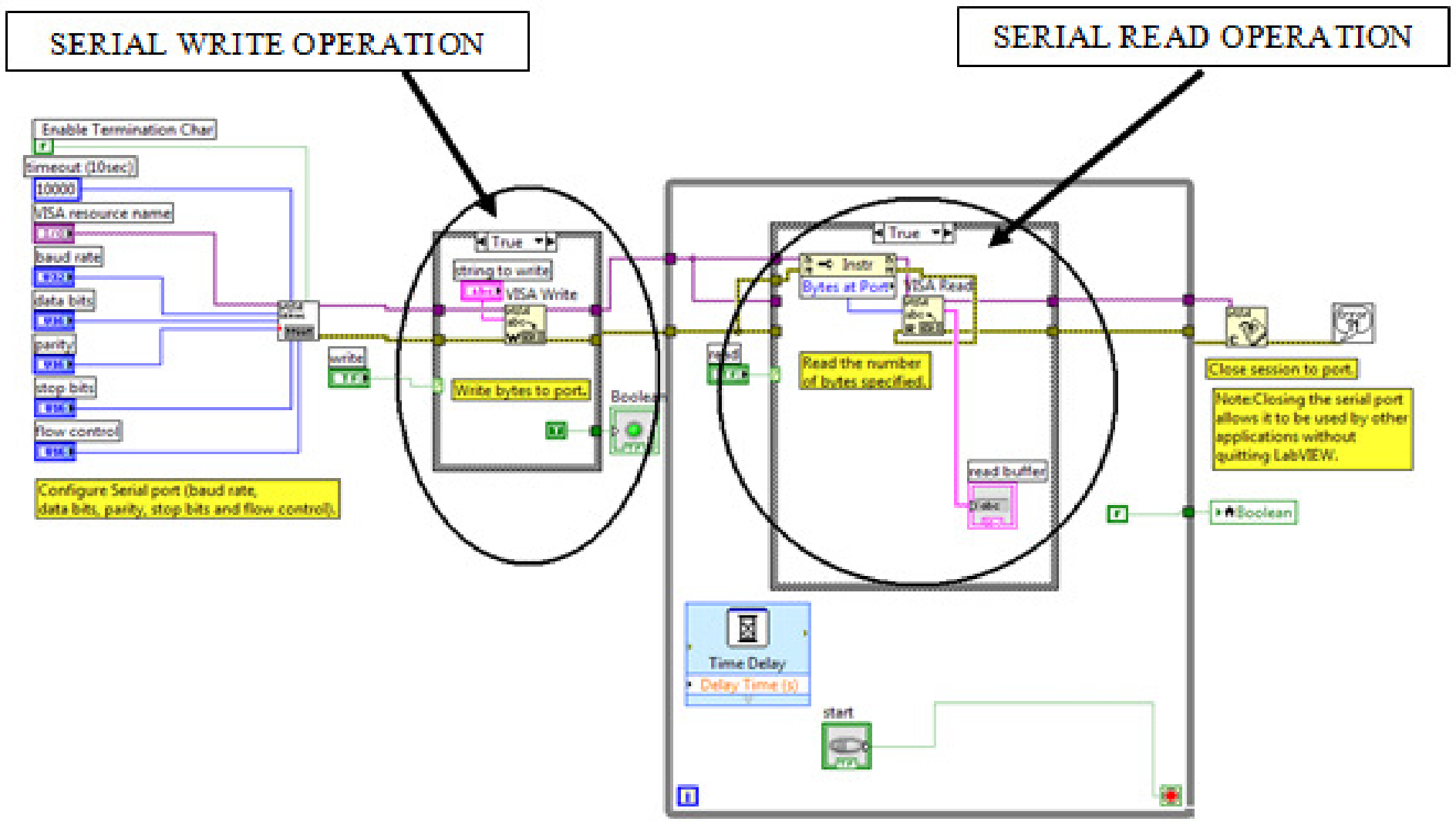

Figure 24. A part of LabVIEW Block Diagram for serial communication using NI-VISA

The LabVIEW Block Diagram implementation for the GUI monitoring the sensor status on the H-track is shown in figure 25.

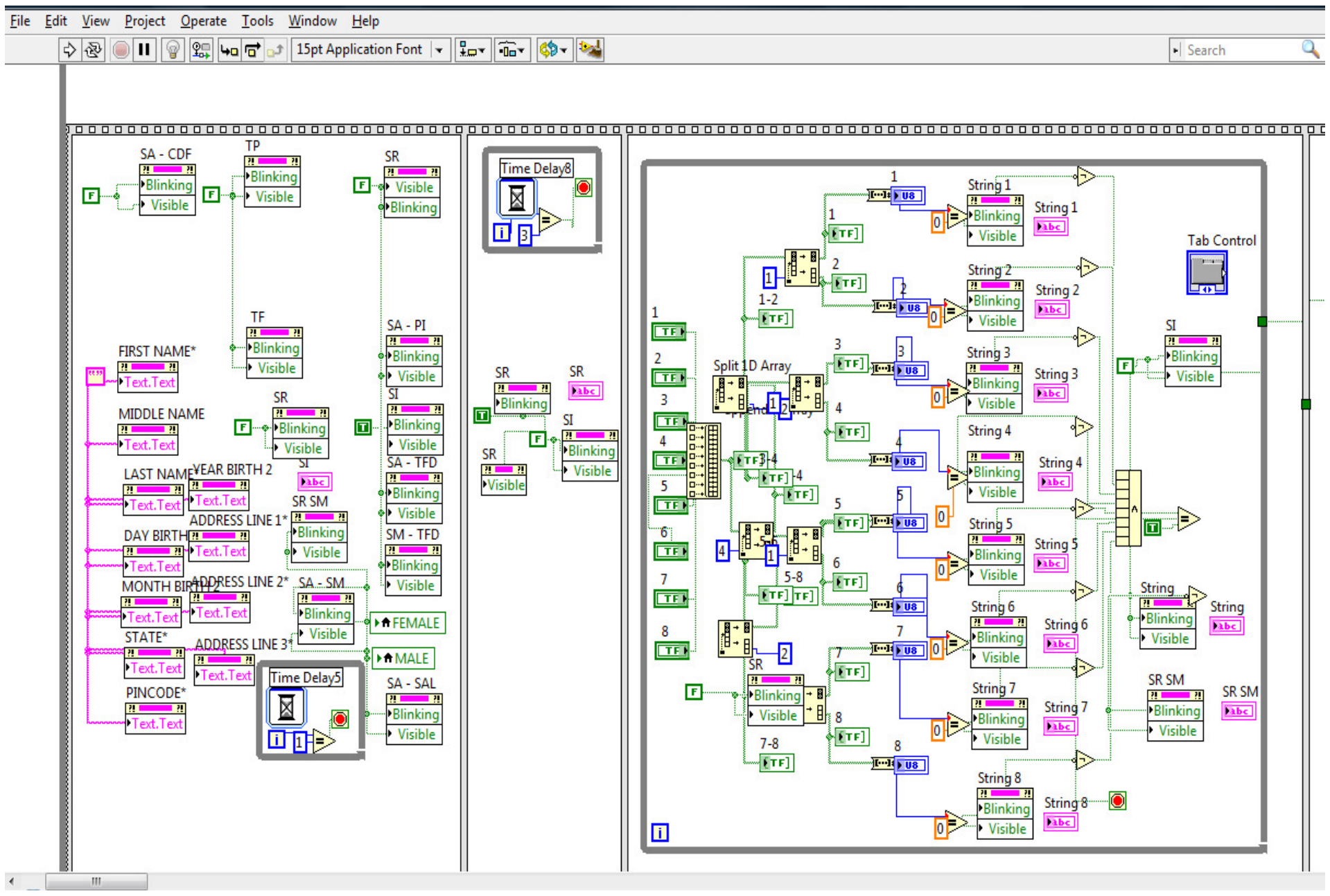

Figure 25. A part of LabVIEW Block Diagram for monitoring sensor embedded H track

The LabVIEW Block Diagram implementation for the e-application GUI is shown in figure 26.

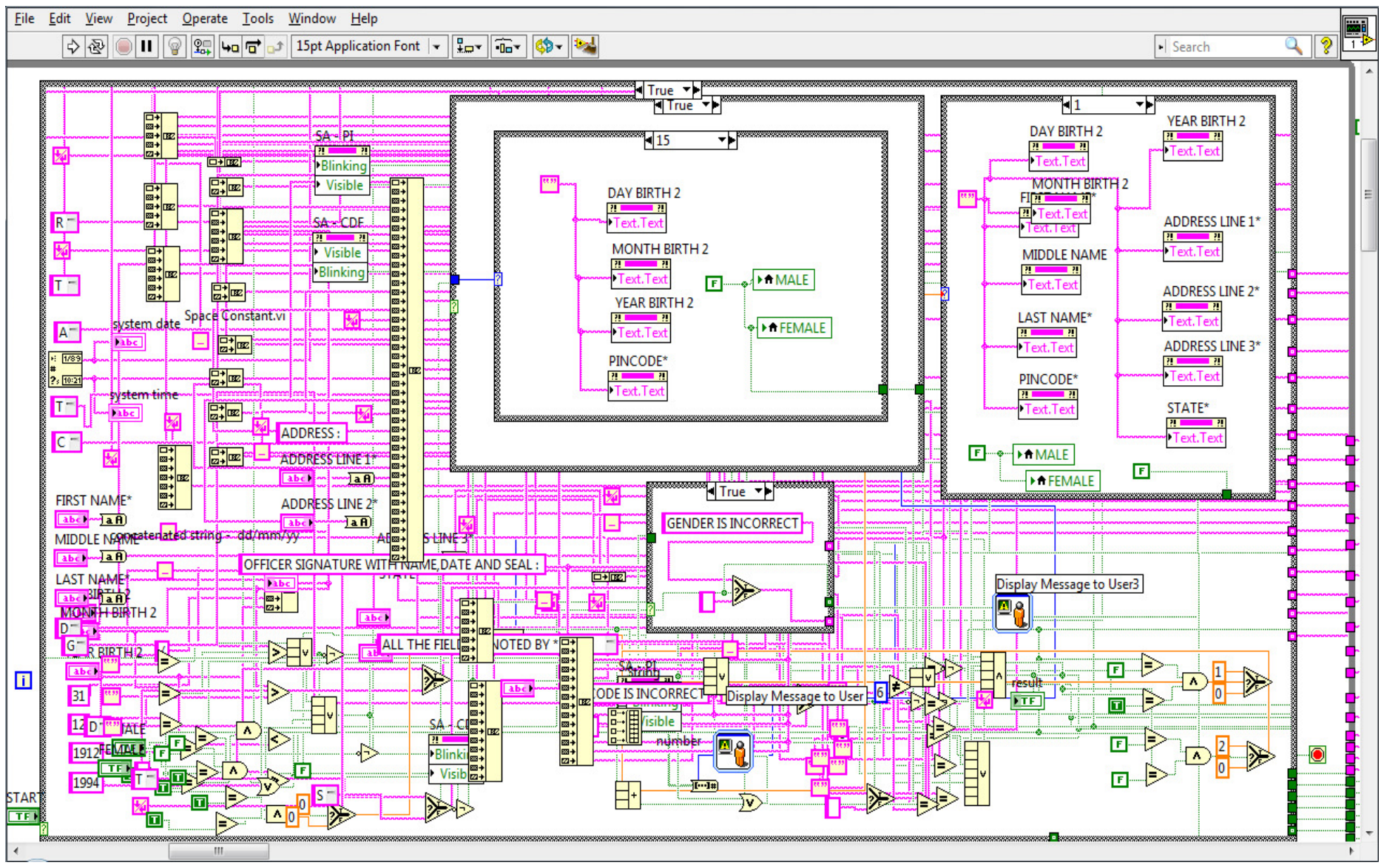

Figure 26. A part of LabVIEW Block Diagram for e-application

## 5. CONCLUSIONS

A Smart Driving Skill Evaluation (SDSE) system using ZigBee based wireless acquisition is discussed. The usage of LabVIEW based technology for skill assessment in the automated driving test process eliminates human intervention leaving no scope for manipulation and negotiation. Hence we can say that the SDSE system increases the level of transparency in the driving skill test process and decreases the rate of corruption in the process of issuing the driving license.

## ACKNOWLEDGEMENTS

The authors would like to thank Mr. Babu Sanker S, Asst. Professor of the Mechanical Engineering department, St. Joseph College of Engineering and Technology, Kerala for his ideas in implementing this work.

**Authors**

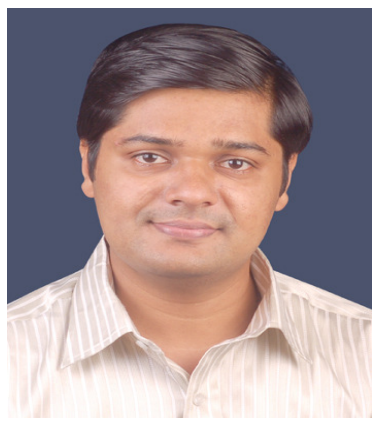

**Mohit John** received his B.Tech degree in Electronics and Instrumentation Engineering from College of Engineering Kidangoor affiliated to Cochin University of Science and Technology Kerala, India and his M.Tech degree in Embedded Systems from DOEACC – Calicut, affiliated to University of Calicut, India. Currently he is working as an Assistant Professor in the department of Applied Electronics and Instrumentation Engineering in St. Joseph College of Engineering and Technology Kerala, India. His research interest includes Embedded System, RTOS, Power Electronics, Automation and Digital System Design.

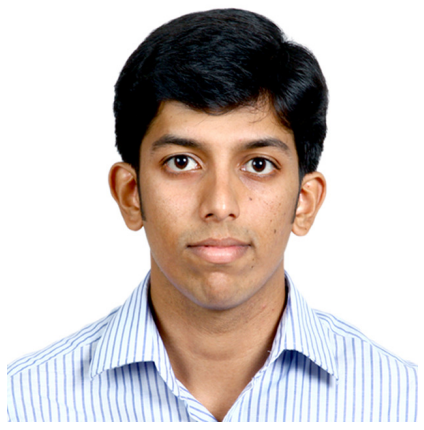

**Arun Joaeph** received his B.Tech degree in Applied Electronics and Instrumentation Engineering from Rajagiri School of Engineering and Technology Kakkanad affiliated to Mahatma Gandhi University Kerala, India and his M.E degree in Control and Instrumentation from Karunya University, India. Currently he is working as an Assistant Professor in the department of Applied Electronics and Instrumentation Engineering in St. Joseph College of Engineering and Technology Kerala, India. His research interest includes Process Control and Instrumentation, Computer Aided Process Control, Data Acquisition, Virtual Instrumentation, Automation and Control System